\documentclass[10pt,journal,compsoc]{IEEEtran}

\ifCLASSOPTIONcompsoc
  \usepackage[nocompress]{cite}
\else
  \usepackage{cite}
\fi

\usepackage{graphicx}
\usepackage{amsmath}
\usepackage[T1]{fontenc}
\usepackage{subfig}
\usepackage{color}
\begin{document}

\title{A Dual Sensor Computational Camera for \\High Quality Dark Videography}

\author{Yuxiao Cheng, Runzhao Yang, Zhihong Zhang, Jinli Suo, and Qionghai Dai 
\thanks{All the authors are with the Department of Automation, Tsinghua University.
Jinli Suo and Qionghai Dai are also affiliated with the Institute of Brain and Cognitive Sciences, Tsinghua University. Emails: chengyx18@mails.tsinghua.edu.cn; yangrz20@mails.tsinghua.edu.cn; 	
zhangzh19@mails.tsinghua.edu.cn; jlsuo@tsinghua.edu.cn; qhdai@tsinghua.edu.cn.}
\thanks{Corresponding authors: Jinli Suo and Qionghai Dai.}
}


\IEEEtitleabstractindextext{%
\begin{abstract}
  Videos captured under low light conditions suffer from severe noise. A variety of efforts have been devoted to image/video noise suppression and made large progress. However, in extremely dark scenarios, extensive photon starvation would hamper precise noise modeling. Instead, developing an imaging system collecting more photons is a more effective way for high-quality video capture under low illuminations. In this paper, we propose to build a dual-sensor camera to additionally collect the photons in NIR wavelength, and make use of the correlation between RGB and near-infrared (NIR) spectrum to perform high-quality reconstruction from noisy dark video pairs. In hardware, we build a compact dual-sensor camera capturing RGB and NIR videos simultaneously. Computationally, we propose a dual-channel multi-frame attention network (DCMAN) utilizing spatial-temporal-spectral priors to reconstruct the low-light RGB and NIR videos. In addition, we build a high-quality paired RGB and NIR video dataset, based on which the approach can be applied to different sensors easily by training the DCMAN model with simulated noisy input following a physical-process-based CMOS noise model. Both experiments on synthetic and real videos validate the performance of this compact dual-sensor camera design and the corresponding reconstruction algorithm in dark videography.
\end{abstract}

}

\maketitle

\IEEEdisplaynontitleabstractindextext

\IEEEpeerreviewmaketitle

\IEEEraisesectionheading{\section{Introduction}\label{sec:introduction}}

\IEEEPARstart{U}{nder} low light conditions, video photographers usually have to set high ISO (sensitivity) and short exposure time to avoid motion blur, and thus suffer from noise deterioration badly. 
To address this issue, researchers have been attempting to develop various low-light image/video reconstruction methods in the past decades. 
The basic idea of noise suppression or removal is introducing prior information of nature scenes to recover the latent clean version, and some methods can achieve excellent reconstruction results, such as \cite{ulyanovDeepImagePrior2018}\cite{clausVidennDeepBlind2019}\cite{tassanoFastdvdnetRealtimeDeep2020}\cite{godardDeepBurstDenoising2018}.
The reconstruction can be conducted on either a single image or a video sequence. Generally speaking, multi-frame noise reduction methods usually get better results than those based on a single frame, thanks to the utilization of priors from temporal redundancy of dynamic scenes. 

\begin{figure}[t!]
  \centering
  \includegraphics[width=\linewidth]{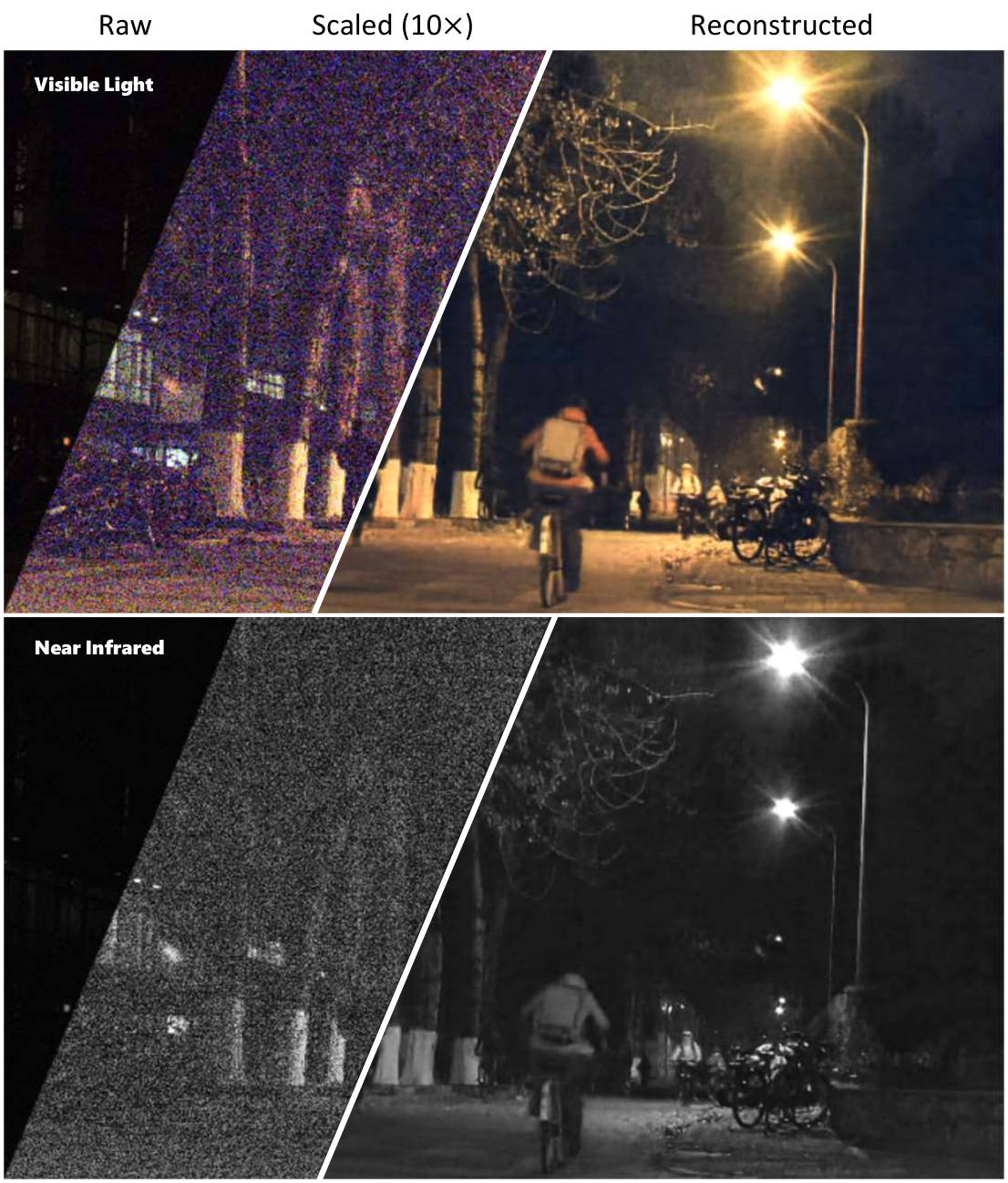}
  \caption{An example of paired RGB (top) and NIR (bottom) video frame captured computationally by our approach under dark environment. The three segments in each image show the extremely dark raw measurement, linearly scaled version with severe noise and color distortion, and the clean final reconstruction, respectively.}
  \label{fig:teaser}
\end{figure}

In spite of the extensive studies in spatial and temporal priors of nature image/videos and their applications in reconstruction tasks, high-quality imaging under extremely low illuminations is still quite challenging. For example, in Fig.~\ref{fig:teaser}, the raw measurement is quite low, and after $\sim$10-time magnification, only the general scene structure can be perceived while the details are buried in severe noise and there exists color distortion. To achieve more decent imaging under environments with poor illuminations, one can resort to collecting more photons and introducing new priors. Actually, there exist a bunch of near-infrared lights under low illuminations and a typical camera sensor can respond to photons from wavelength from 350-1200nm \cite{krishnanDarkFlashPhotography2009}. Therefore, recording the photons in NIR band additionally can improve the quality of dark imaging. However, existing approaches are lacking in the following aspects: (i) An infrared cut-off filter is placed before the image sensor of RGB/gray camera to cut off NIR light for a natural tone matching the perception of human eyes (RGB, roughly 400-700nm) \cite{jiangMultiSpectralRGBNIRImage2019}. As a result, severe noise occurs due to the insufficient photons in visible wavelength. (ii) RGB and NIR images share similar features on scene structure and relative brightness, as shown in the two rows in Fig.~\ref{fig:teaser}, and it is possible to make use of such cross-channel correlations between RGB and NIR images to enhance the imaging quality algorithmically. However, current denoising methods rarely utilize this cross channel prior.

\begin{figure}[t!]
  \centering
  \includegraphics[width=1.64in]{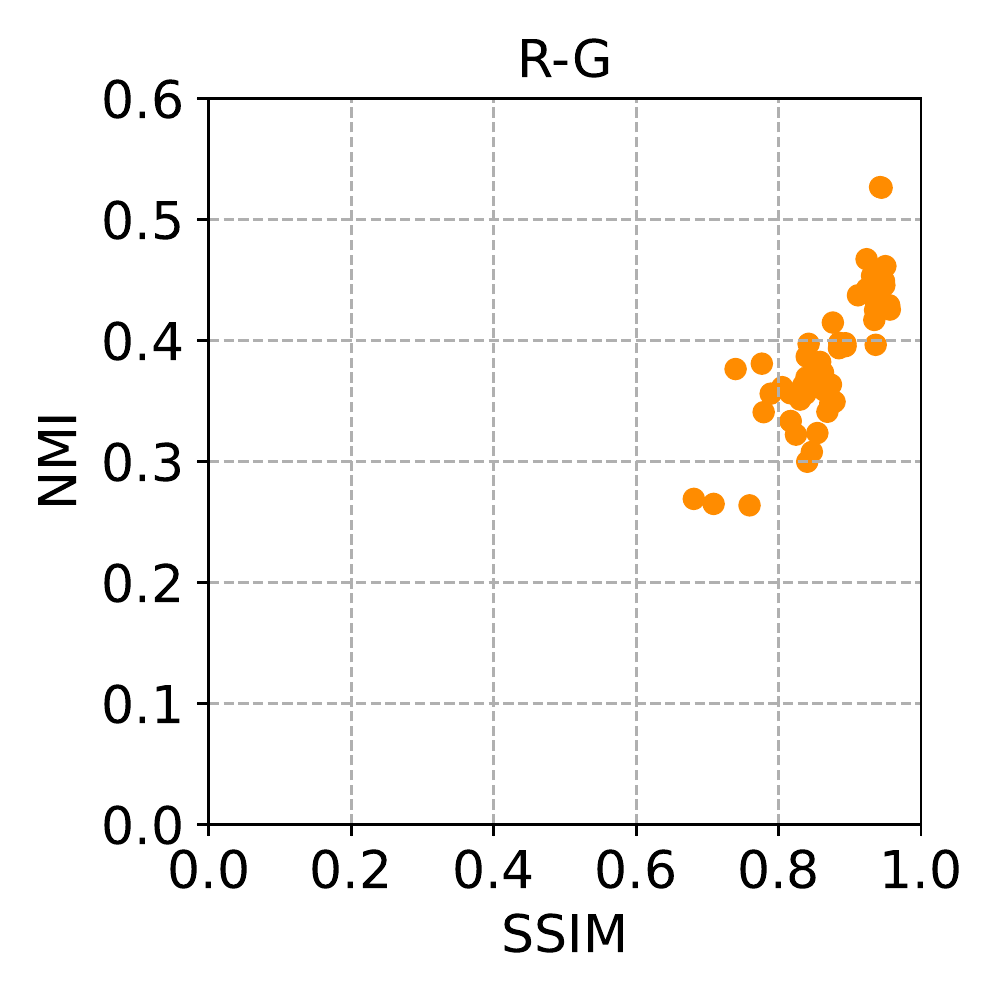}
  \includegraphics[width=1.64in]{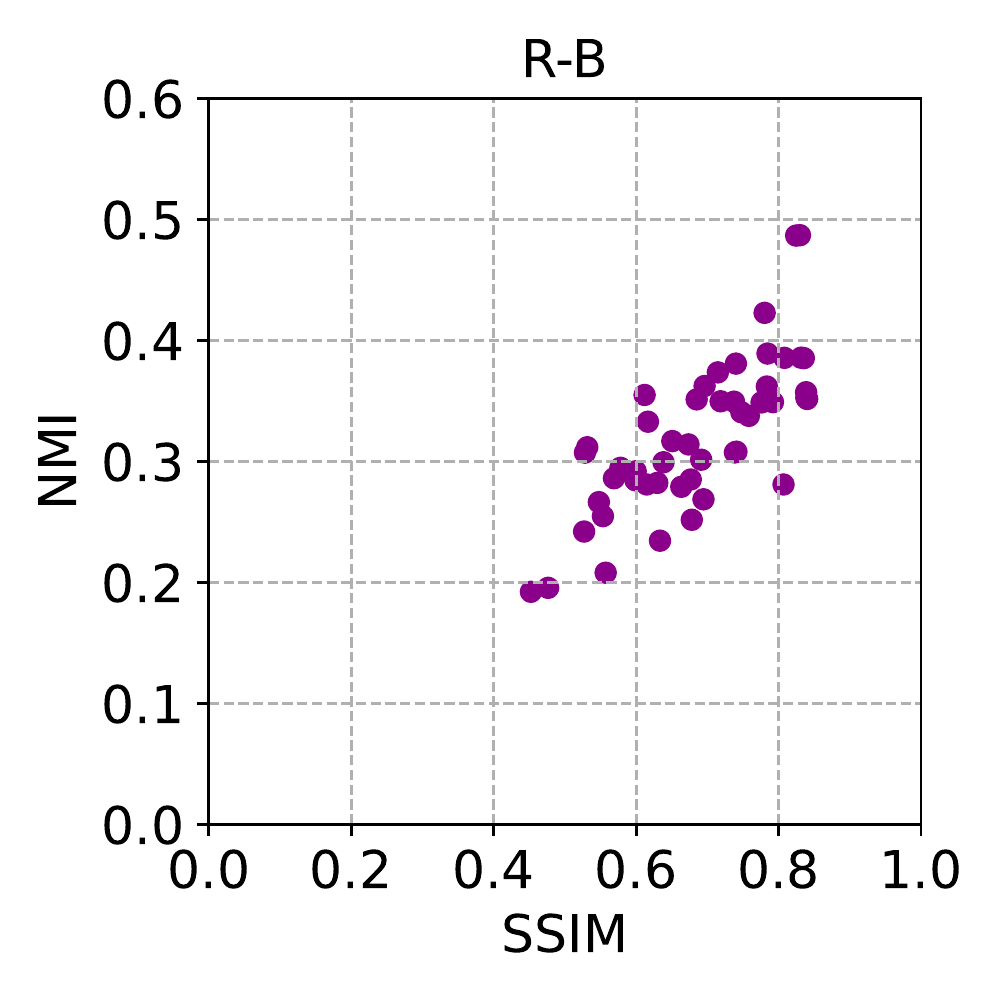}
  \includegraphics[width=1.64in]{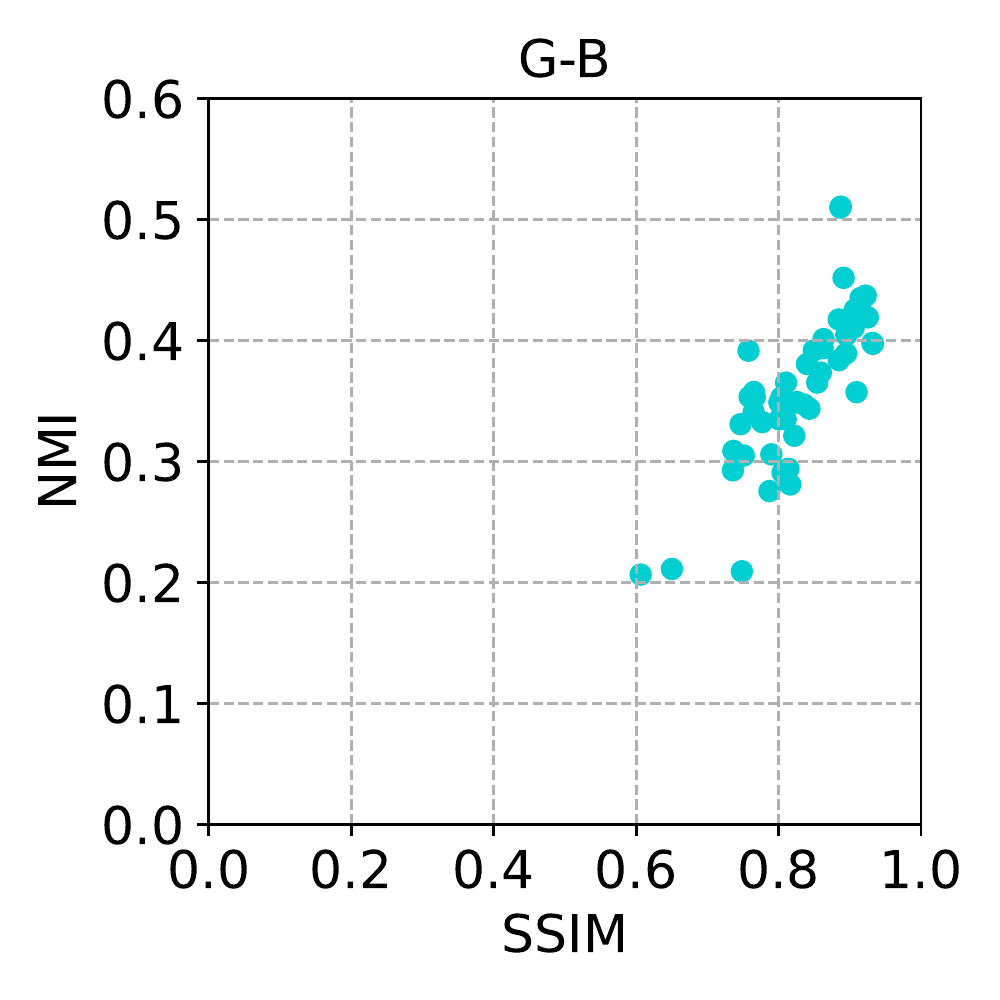}
  \includegraphics[width=1.64in]{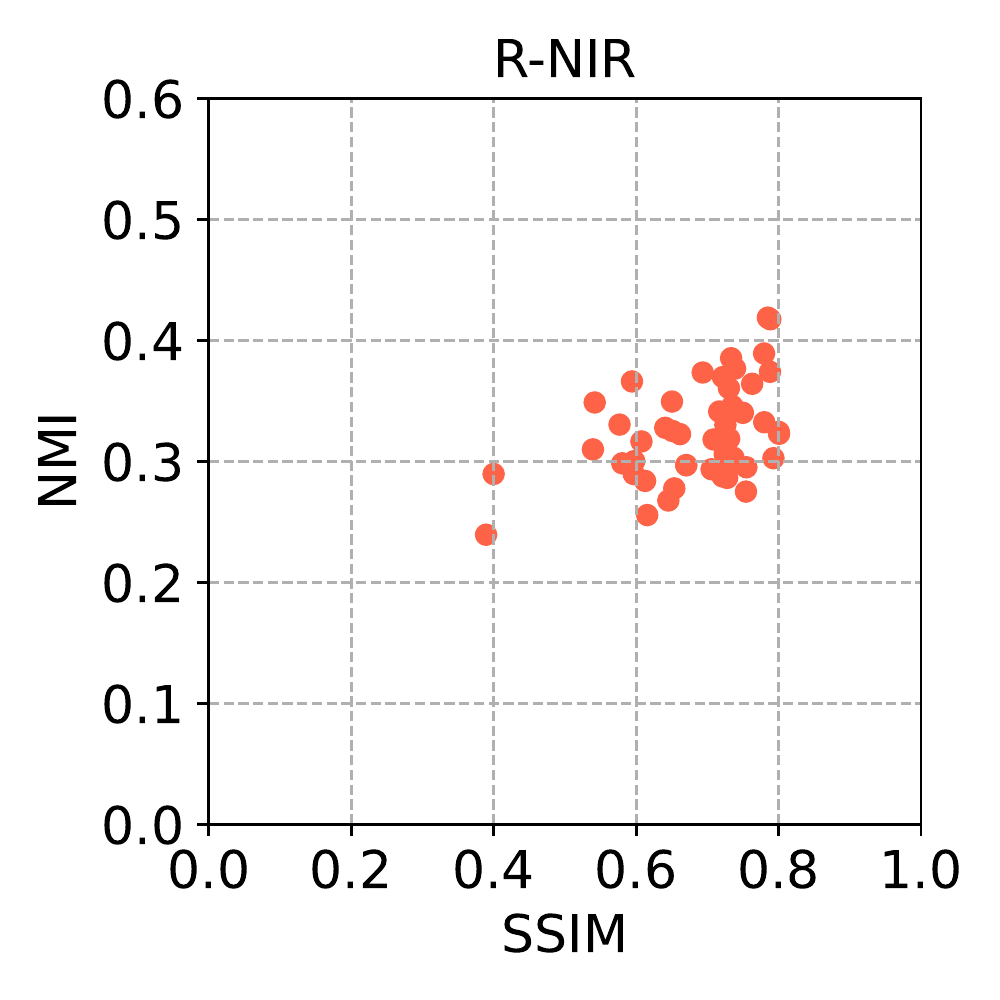}
  \includegraphics[width=1.64in]{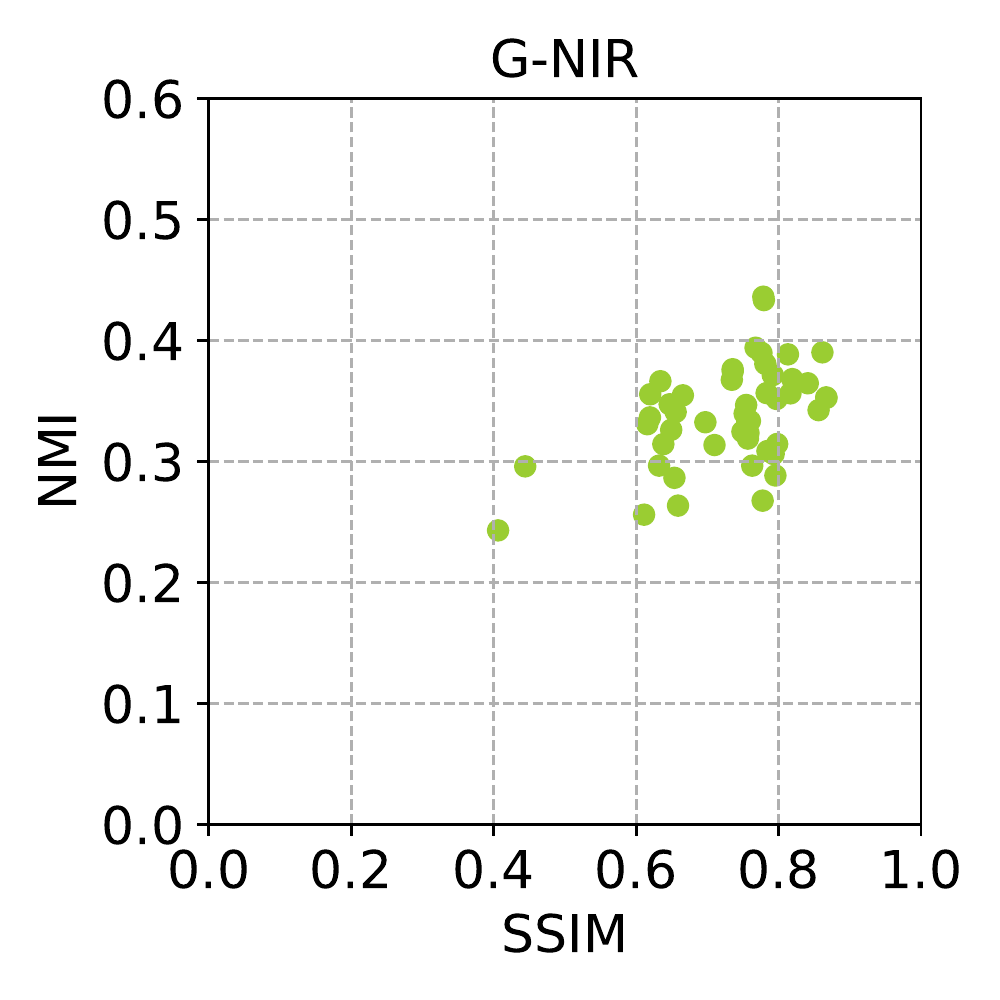}
  \includegraphics[width=1.64in]{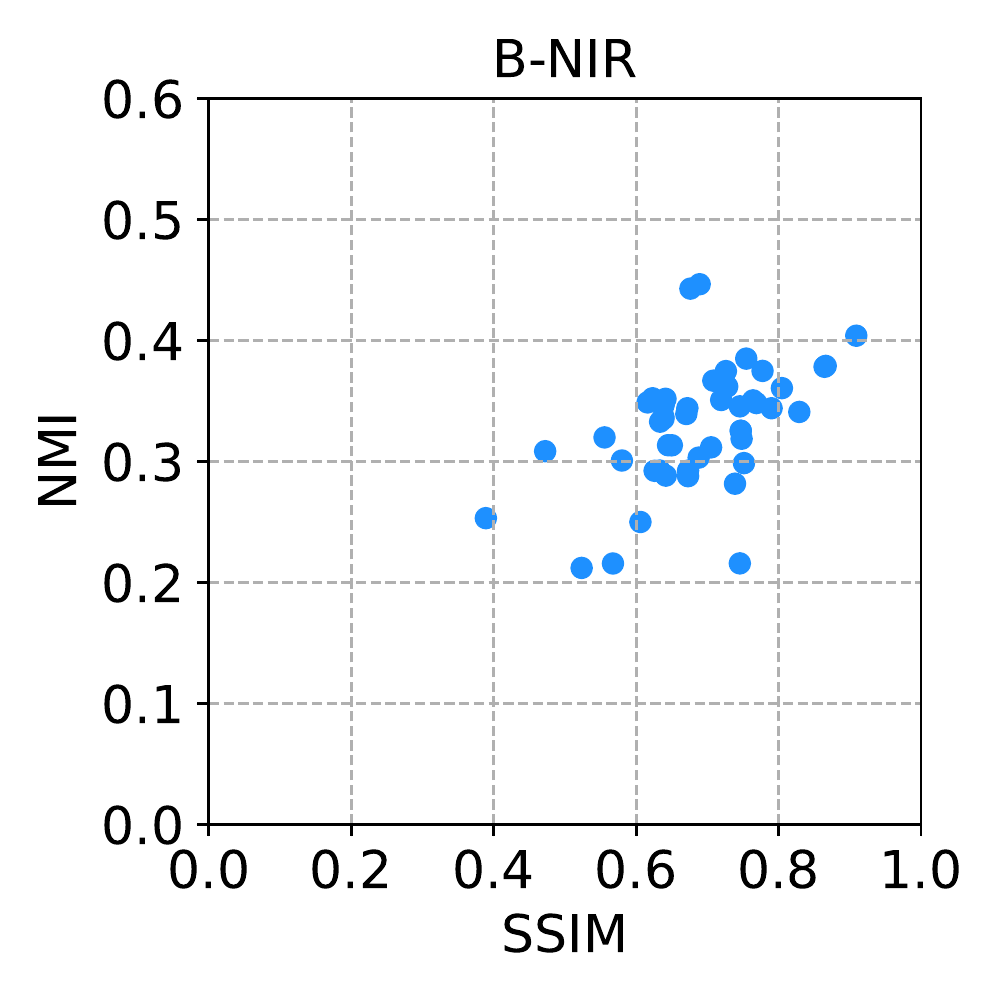}
  \caption{Analysis of the cross channel correlation among three visible channels (i.e., red, green, blue), and between RGB and NIR channels, with SSIM and NMI  as score metrics. These plots are calculated from 50 RGB/NIR image pairs.}
  \label{fig:plot_corr}
  \vspace{-2mm}
\end{figure}

Inspired by the above analysis, we design a compact double-sensor camera system to acquire paired RGB and NIR images at the same time, with a beam splitter placed between the primary lens and image sensors. 
Using this setup, we capture a dataset of RGB-NIR nature video pairs, and analyze the image similarity among the 4 channels (R, G, B, and NIR), in terms of SSIM (Structural Similarity) and NMI (Normalized Mutual Information) scores. The results are plotted in Fig.~\ref{fig:plot_corr} with the scattered dots concentrated in the right-top region of each subfigure, which validates that the red, green, blue, and infrared channels are indeed strongly correlated, and the correlations between visible and NIR wavelength are only slightly lower than that among R, G, and B channels. In other words, we can introduce priors across visible and NIR channels to improve the final reconstruction.


Computationally, we design a Dual-channel Multi-frame Attention Network (DCMAN) based on bi-directional ConvLSTM and CNN-based encoder-decoder with mutual multi-level feature fusion \cite{zhangHyperfusionNetHyperdenselyReflective2019}. In other words, the data from the two arms contribute mutually to recovering the latent structure of nature videos buried in the severe noise. By additionally introducing the photons in the near-infrared wavelength, and taking advantage of spatial-temporal as well as spectral prior information, we can achieve better reconstruction quality. Our method takes noisy NIR and RGB videos as inputs and gives clean results of both videos at the same time, as shown in Fig.~\ref{fig:teaser}. The result demonstrates that the visual quality is largely raised with fine details being reconstructed.

Overall, this paper  reports a computational photography approach for dark video capturing and mainly contributes in following aspects:

\begin{itemize}
  \item We build a dual-sensor camera system capturing RGB and NIR video pairs simultaneously, being able to increase the collected photons effectively for higher imaging quality in dark environments. 
  \item We design a Dual-Channel Multi-frame Attention Network (DCMAN) with Guided Skip Connections (GSCs) to utilize spatial, temporal, and spectral video prior. This network can serve as a general structure applicable for similar tasks taking multiple input channels.
  \item The approach can be adapted to different sensors and noise levels easily by learning the model from synthetic data, produced by adding noise to our collected high quality RGB + NIR video dataset on real scenes.
  \item The proposed approach can achieve superior performance to existing methods in most dark scenarios.
\end{itemize}

\section{Related Works}


\subsection{Video Denoising}
Video denoising has been extensively studied in the past decades \cite{zhangBrighteningLowlightImages2021}\cite{tianDeepLearningImage2018}\cite{s.rNoiseReductionVideo2012}, and the large amount of literature falls into several groups based on the ways of specifying priors. 

Statistically, the distribution of natural images/videos is discriminative from noises globally or over local patches. One way of noise removal is applying filters spatiotemporally or in a transformed domain after motion compensation, making use of the fact that noise components are of high frequency while the latent image/video concentrates at low frequency band \cite{jinWaveletVideoDenoising2006}\cite{zlokolicaWaveletdomainVideoDenoising2006}\cite{braileanNoiseReductionFilters1995}. 
Another way is conducting video denoising over patches, exploiting temporal and spatial patch similarity in the neighborhood. For example, Buates et al. \cite{buadesPatchbasedVideoDenoising2016} propose to conduct patch-based denoising after compensating motions among neighboring frames, and some other methods can suppress noise without motion compensation, such as non-local means algorithm \cite{buadesDenoisingImageSequences2005},  and V-BM3D \cite{dabovVideoDenoisingSparse03} and its variants. 
V-BM3D \cite{dabovVideoDenoisingSparse03} is the extension of patch-based method BM3D \cite{lebrunAnalysisImplementationBM3D2012}, which extracts similar 2D patches from consecutive video frames and then stacks them for filtering in a transformed domain. Later, they extend V-BM3D to V-BM4D \cite{maggioniVideoDenoisingDeblocking2012} searching similar 3D spatio-temporal blocks instead of 2D patches, and VNLB \cite{ariasVideoDenoisingEmpirical2018} that conducts inference with Bayesian estimation. 

With the rapid development of Convolutional Neural Networks (CNN) and deep learning techniques, learning-based image reconstruction exhibits better performance and is attracting wide attention. In this stream, video denoising, however, is much less explored than image denoising, since extracting spatio-temporal features jointly is non-trivial. On the one hand, researchers try to extend patch-based image denoising by introducing temporal information with CNN feature fusion. Tassano et al. \cite{tassanoDvdnetFastNetwork2019} propose DVDnet by explicitly splitting the denoising process into two successive stages. Specifically, consecutive frames are first individually denoised and then temporally denoised with motion compensation. Later, building on DVDnet, they further propose FastDVDnet \cite{tassanoFastdvdnetRealtimeDeep2020} without flow estimation and achieved higher performance. Using a similar two-stage method, Claus and Gemert \cite{clausVidennDeepBlind2019} design ViDeNN, and Mildenhall et al. \cite{mildenhallBurstDenoisingKernel2018} design a Kernel Prediction Network (KPN) to suppress noises in a burst of dark images.  
On the other hand, some researchers explore Recurrent Neural Networks for extracting temporal information. Chen et al. \cite{chenDeepRnnsVideo2016} develop an RNN-based method to utilize temporal video prior for noise reduction. Wang et al. \cite{wangEnhancingLowLight2019} introduce an LSTM-based method, extracting both short and long-term dependencies from image sequences. Godard et al. \cite{godardDeepBurstDenoising2018} propose a global recurrent network and append it to existing CNN-based single image denoising networks. 
Beyond these works on spatio-temporal feature extraction, we explore the possibility of making use of spatial, temporal, as well as spectral corrections in one end-to-end network---CNN for spatial information, LSTM for temporal information, and finally network channel fusion for spectral RGB-NIR information.

\subsection{Noise Modeling}
The noise in low light photography is complex and closely correlated with camera sensors. A precise description of the sensor noise is crucial for the final denoising performance. In the early stage, most of the denoising methods assumes identically distributed (i.i.d.) additive white Gaussian noise (AWGN) and are of limited performance. Recently, several new noise models are explored, such as mixture of Gaussian (MoG) \cite{zhuNoiseModelingBlind2016}, Poisson-Gaussian \cite{makitaloNoiseParameterMismatch2014} and other physical-process-based mixture model \cite{healeyRadiometricCCDCamera1994}\cite{tsinStatisticalCalibrationCCD2001}\cite{weiPhysicsbasedNoiseFormation2020}\cite{yueSupervisedRawVideo2020}. Since high sensitivity cameras are preferred in low light imaging, Wang et al. \cite{wangEnhancingLowLight2019} model CMOS noise with a more complex high sensitivity noise model and propose a noise calibration method assisting generating plausible synthetic noisy images/videos.
In this paper, we build an integrated
physical-process-based noise model covering rich noise sources, with high performance and good compatibility with different sensors. 

\subsection{RGB-NIR Imaging}
The short-wavelength NIR images are of similar brightness and structures to RGB images.
Especially, under low-light conditions NIR images serve good performance \cite{hanInfraredImageSuperresolution2018} and exhibit much more details invisible to eyes \cite{chenRGBNIRMultispectralCamera2014}, so introducing NIR to RGB cameras can be a promising way for low light photography.
Fortunately, widely used CCD and CMOS can detect photons in short-wavelength NIR range up to 1200 nm, which largely facilitate RGB-NIR hybrid imaging. 
Towards this direction, researchers have made some progress. Some groups use NIR images to assist with high-quality RGB imaging. For example, Krishnan and Fergus \cite{krishnanDarkFlashPhotography2009} and Zhuo et al. \cite{zhuoEnhancingLowLight2010} 
enhance RGB images captured with ambient light using images captured with NIR flashlight (dark flash), whereas the former captures two images sequentially and the latter uses two synchronized hybrid cameras to take the pair simultaneously. Similarly, Sugimura et al. \cite{sugimuraEnhancingColorImages2015} construct an imaging system to take RGB and NIR photos with different exposure times and recover clean color images from those image pairs. Conversely, Han et al. \cite{hanInfraredImageSuperresolution2018} propose to super-resolve low-resolution NIR images using a high-resolution RGB image. There are also works trying to capture both high-resolution NIR and RGB images computationally, for example, Hu et al. \cite{huConvolutionalSparseCoding2018} propose RGB-NIR reconstruction techniques in a single sensor with a novel RGB-NIR Color Filter Array (CFA). In spite of the above progress, there is still no working can capture and enhance both RGB and NIR for dark visual recording.

In this paper, we build a compact RGB/NIR imaging system to capture paired videos, which is largely different from \cite{sugimuraEnhancingColorImages2015} that uses two separate cameras. Such a design is of high stability and can be used in the cases requiring lightweight imaging setups. The reconstruction is also different from the above hybrid imaging by enhancing both RGB and NIR videos to maximize the mutual assistance, instead of focusing on only one of them. In addition, a high-quality RGB-NIR dataset is important for data-driven algorithms, but several existing datasets, such as \cite{niuVIPLHRMultimodalDatabase2018} and \cite{brownMultispectralSIFTScene2011}, either consists of only single images (i.e., static scenes) or are constrained to a specific type of scenes. To address these problems, we capture a dataset with paired RGB and NIR nature videos.

\subsection{Channel-Fusion Learning}
Channel fusion is an effective way of taking multiple input data and learning the latent structure from both intra- and inter- multiple data sources and can incorporate CNN-based deep networks. Specifically, the informative feature extracted by CNN can be fused at the input level, early level, or late level \cite{jiangMultiSpectralRGBNIRImage2019}\cite{caballeroRealtimeVideoSuperresolution2017}. However, to leverage features at different levels, researchers propose networks with feature fusion across multiple levels \cite{zhangHyperfusionNetHyperdenselyReflective2019}. The advantages of such multi-level fusion have been successfully applied in high-level vision tasks such as object or scene change detection \cite{zhangHyperfusionNetHyperdenselyReflective2019} and \cite{leiHierarchicalPairedChannel2021}, but the studies are still at an early stage in the field of image reconstruction. 

\begin{figure*}[t]
  \centering
  \includegraphics[width=\textwidth]{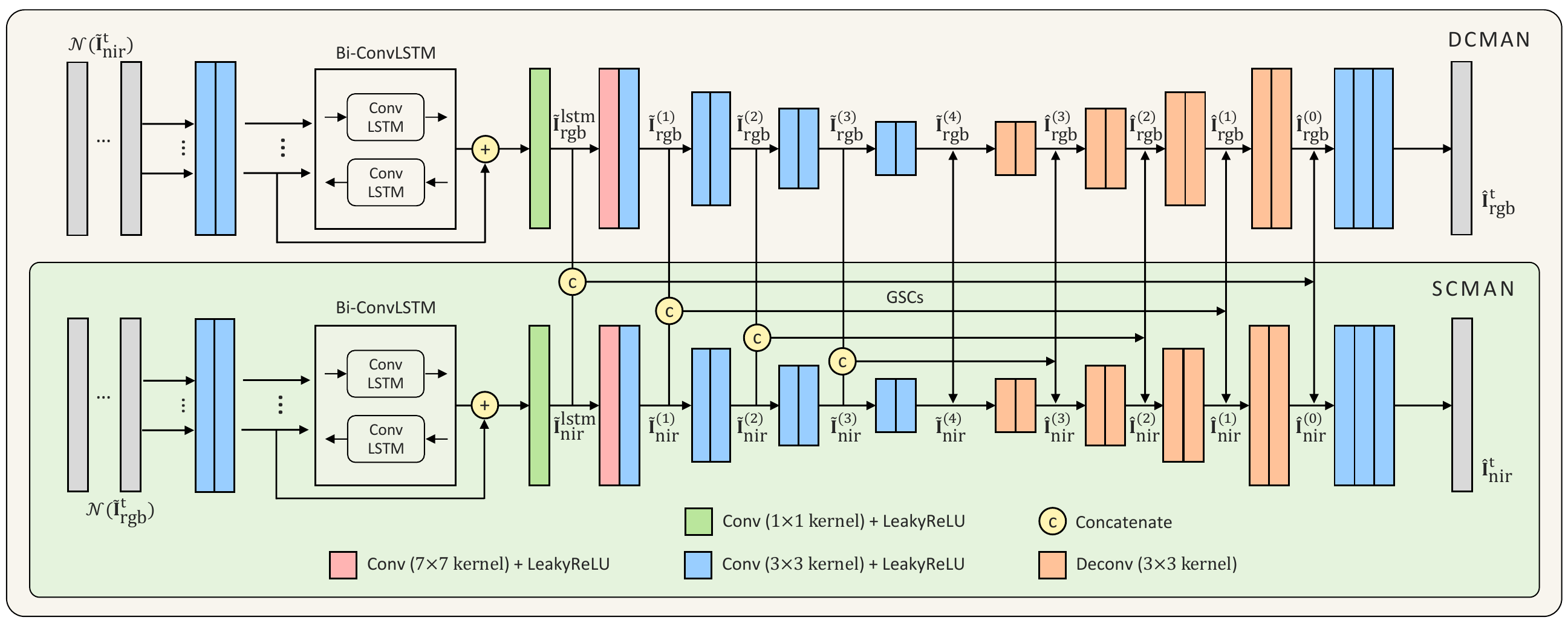}
  \caption{The structure of our proposed Dual-Channel Multi-Frame Attention Network (DCMAN). Our network includes two channels: each channel consists of a ConvLSTM module and a U-Net-shaped CNN module; two channels are fused with Guided Skip Connections (GSCs). This network explores the spatial prior in nature videos with the U-Net CNN module, temporal prior with the Bi-ConvLSTM module, and spectral prior with inter-channel GSCs.}
  \label{net}
\end{figure*}

Meng et al. \cite{mengRobustQualityEnhancement2021} use consecutive frames as dominant data channel while guided maps serve as an assistant arm, which guides the network to focus more on certain regions. Like in \cite{zhangHyperfusionNetHyperdenselyReflective2019}, these two channels are fused at multiple levels to achieve high performance. Similarly, Zha et al. \cite{zhaDetectingOvershootingCloud2020} design a dual-channel multi-scale deep network to fuse the features from multi-spectral bands. Inspired by that work, we can utilize RGB and NIR correlations explicitly. In this paper, the proposed dual-channel multi-frame attention network (DCMAN) is a symmetrical structure, which treats two channels---RGB and NIR---in a parallel manner to mutually enhance each other's quality, considering the similar photon counts and noise levels at two corresponding input light paths.

\section{Network Structure}

We propose a Dual-Channel Multi-frame Attention Network (DCMAN) with two channels taking noisy RGB and NIR inputs: each channel is equipped with a ConvLSTM module and a U-Net-shaped CNN module to explore the spatio-temporal redundancies for noise suppression; two channels are fused with Guided Skip Connections at multiple layers to explore cross-spectrum prior in nature videos. 

The training dataset is composed of paired noisy and clean video patches cropped at the same location in consecutive $2T+1$ frames, each entry can be described as
\begin{equation}
\left\{ <\mathbf{I}^{1\cdots (2T+1)}_{\text{rgb}} , \mathbf{I}^{1\cdots (2T+1)}_{\text{nir}}>,  <\mathbf{\widetilde{I}}^{1\cdots (2T+1)}_{\text{rgb}},  \mathbf{ \widetilde{I}}^{1\cdots (2T+1)}_{\text{nir}}>\right\}
\end{equation}
Here $\mathbf{I}^t_{\text{rgb}}$ and $\mathbf{I}^t_{\text{nir}}$ denote a clean RGB and corresponding NIR patch captured under good illumination, $\mathbf{\widetilde{I}}^t_{\text{rgb}}$ and $\mathbf{\widetilde{I}}^t_{\text{nir}}$ denote noisy counterparts generated following the physical based noise model, and $t$ denotes the frame index. We define the neighboring frames of $\mathbf{I}^t$ as
\begin{equation}
    \mathcal{N}(\mathbf{I}^t) = \{ \mathbf{I}^{t-T},\mathbf{I}^{t-T+1}, \cdots \mathbf{I}^{t-1},\mathbf{I}^{t+1},\cdots \mathbf{I}^{t+T-1}, \mathbf{I}^{t+T}\}.
\end{equation}
In the next subsections, we describe the processing of $t$th frame, and omitted the subscript $t$ for simplicity.

\subsection{Bi-ConvLSTM Subnet}
Utilizing the redundancies among neighboring video frames, the noise deterioration in a video frame can be corrected by neighboring frame patches, including past and future ones. To explore the latent temporal correlation, we first equip the network with two identical bi-directional ConvLSTM \cite{mengRobustQualityEnhancement2021}\cite{xingjianConvolutionalLSTMNetwork2015} modules, which process the input RGB patch sequence $\mathcal{N}(\mathbf{\widetilde{I}}_{\text{rgb}})$ and NIR patches $\mathcal{N}(\mathbf{\widetilde{I}}_{\text{nir}})$ separately after an affiliated pre-convolutional layers. The residual connections in bi-directional ConvLSTM (bi-ConvLSTM) will guide the network to penalize large differences between adjacent frame patches. Specifically, the bi-ConvLSTM $F_{\text{lstm}}\left(\cdot\right)$ can be formulated as follows:
\begin{eqnarray}
\mathbf{\widetilde{I}}^{\text{lstm}}_{\text{rgb}} &= F_{\text{lstm}}\left( \mathcal{N}(\mathbf{\widetilde{I}}_{\text{rgb}}); \theta^{\text{lstm}}_{\text{rgb}} \right) \label{lstm1} \\
\mathbf{\widetilde{I}}^{\text{lstm}}_{\text{nir}} &= F_{\text{lstm}}\left(  \mathcal{N}(\mathbf{\widetilde{I}}_{\text{nir}}); \theta^{\text{lstm}}_{\text{nir}} \right) \label{lstm2},
\end{eqnarray}
with  $\theta^{\text{lstm}}_{\text{rgb}}$ and $\theta^{\text{lstm}}_{\text{nir}}$ being the network parameters of $F_{\text{lstm}}\left(\cdot\right)$ to be learned.
The receptive field of this module depends on the number of pre-Conv and bi-ConvLSTM layers. Usually the motion between adjacent frames is relatively mild, a shallow bi-ConvLSTM module would be sufficient.

\subsection{Dual-Channel Encoder}

To further extract the spatio-temporal-spectral information, we design a Dual-Channel subnet based on encoder-decoder networks. The encoder module of each channel has 4 encode layers, each layer takes output of the the former layer as input. The input of the first layer $\mathbf{\widetilde{I}}^{(0)}_{\text{rgb}}$ and $\mathbf{\widetilde{I}}^{(0)}_{\text{nir}}$ are the output of bi-ConvLSTM module, i.e., $\mathbf{\widetilde{I}}^{(0)}_{\text{rgb}} = \mathbf{\widetilde{I}}^{\text{lstm}}_{\text{rgb}}$ and $ \mathbf{\widetilde{I}}^{(0)}_{\text{nir}} = \mathbf{\widetilde{I}}^{\text{lstm}}_{\text{nir}}$. The encoder can be formulated as
\begin{eqnarray}
\mathbf{\widetilde{I}}_{\text{rgb}}^{(k)} &=& F_\downarrow^{(k)}\left(\mathbf{\widetilde{I}}^{(k-1)}_{\text{rgb}} ; \theta_{r\downarrow}^{(k)}\right) \\
\mathbf{\widetilde{I}}_{\text{nir}}^{(k)} &=& F_\downarrow^{(k)}\left(\mathbf{\widetilde{I}}^{(k-1)}_{\text{nir}} ; \theta_{n\downarrow}^{(k)}\right),\ k \in [1, 4].
\end{eqnarray}
In these two equations, $F_\downarrow^{(k)}$ is the encoder layer at the $k$th level with $k\in [1,4]$, $\theta_{\text{rgb}\downarrow}^{(k)} $ and $\theta_{\text{nir}\downarrow}^{(k)}$ are parameters of subnets in two channels, repsectively.

\subsection{Guided Skip Connections in Decoder}
The RGB and NIR channels guide each other mutually with the help of the Guided Skip Connections (GSCs), an extension to Skip Connections \cite{ronnebergerUnetConvolutionalNetworks2015}. GSCs in decoder module can not only synthesize high-resolution low-level features, like the skip connections does, but can also combine features from another channel (RGB or NIR). In our GSCs, the features from encoder are concatenated and passed to the 4-layer decoder  $F_\uparrow^{(k)}$ with $k\in [1,4]$, specifically defined as
\begin{eqnarray}
\mathbf{\hat{I}}_{\text{rgb}}^{(k-1)} &=& F_\uparrow^{(k)}\left( \left\{ \mathbf{\hat{I}}_{\text{rgb}}^{(k)} , \mathbf{\widetilde{I}}_{\text{rgb}}^{(k)}, \mathbf{\widetilde{I}}_{\text{nir}}^{(k)}\right\} \ ;\ \theta_{\text{rgb}\uparrow}^{(k)}\right) \\
\mathbf{\hat{I}}_{\text{nir}}^{(k-1)} &=& F_\uparrow^{(k)}\left( \left\{ \mathbf{\hat{I}}_{\text{nir}}^{(k)} , \mathbf{\widetilde{I}}_{\text{nir}}^{(k)}, \mathbf{\widetilde{I}}_{\text{rgb}}^{(k)}\right\} \ ;\ \theta_{\text{nir}\uparrow}^{(k)}\right).
\end{eqnarray}
Here $\mathbf{\hat{I}}_{\text{rgb}}^{(4)} = \mathbf{\widetilde{I}_{\text{rgb}}^{(4)}},\ \mathbf{\hat{I}_{\text{nir}}^{(4)}} = \mathbf{\widetilde{I}}_{\text{nir}}^{(4)}$, i.e., output of the encoder module. The final output of our network are 

\begin{eqnarray}
  \mathbf{\hat{I}}_{\text{rgb}} &=& F^{(0)}\left( \left\{ \mathbf{\hat{I}}_{\text{rgb}}^{(0)} , \mathbf{\widetilde{I}}_{\text{rgb}}^{(0)}, \mathbf{\widetilde{I}}_{\text{nir}}^{(0)}\right\} \ ;\ \theta_{\text{rgb}}^{(0)}\right) \\
  \mathbf{\hat{I}}_{\text{nir}} &=& F^{(0)}\left( \left\{ \mathbf{\hat{I}}_{\text{nir}}^{(0)} , \mathbf{\widetilde{I}}_{\text{nir}}^{(0)}, \mathbf{\widetilde{I}}_{\text{rgb}}^{(0)}\right\} \ ;\ \theta_{\text{nir}}^{(0)}\right)
\end{eqnarray}

\subsection{Loss Function}

The loss function for our network is defined to favor better image sharpness and higher color fidelity jointly. Specifically, 
 the loss function can be written as
\begin{align}
  \begin{aligned}
    L = &\sum_{i} \lambda_1 \left( \left\| \mathbf{\hat{I}}_{i,\text{rgb}} - \mathbf{I}_{i,\text{rgb}} \right\|_1 + \left\| \mathbf{\hat{I}}_{i,\text{nir}} - \mathbf{I}_{i,\text{nir}} \right\|_1 \right) +\\
    & \sum_{i} \lambda_2 L_{\text{cos}}(\mathbf{\hat{I}}_{i,\text{rgb}}, \mathbf{I}_{i,\text{rgb}}), 
  \end{aligned}
\end{align}
in which $i$ indexes patches in the training batch, and $L_{\text{cos}}$ denotes cosine embedding loss\cite{wilkinsonSemanticVerbatimWord2016}
\begin{align}
  \begin{aligned}
  L_{\text{cos}}(\mathbf{x},\mathbf{y}) = 1 -\cos (\mathbf{x},\mathbf{y}).
    \end{aligned}
\end{align}
  
\section{Integrated CMOS Noise Model}

\begin{figure}[t!]
  \centering
  \includegraphics[width=\linewidth]{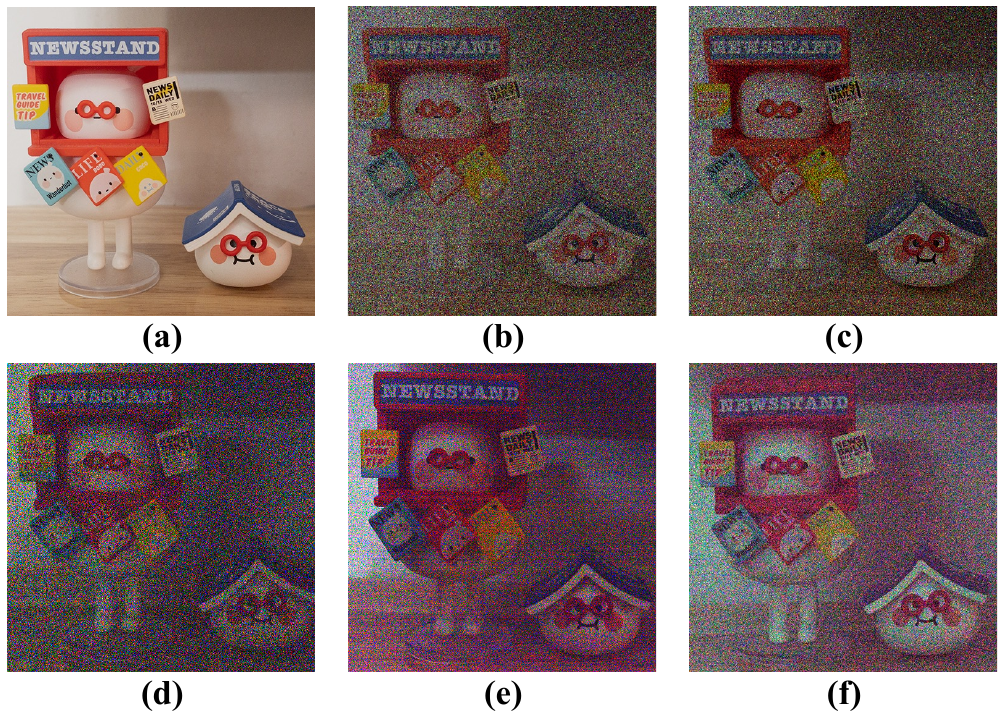}
  \caption{Comparison among the synthetic noisy images of the scene in (a) following different noise models (b-c) and real noisy images captured by different cameras (d-f). (a) Bright image captured under long exposure. (b)(c) Images superimposed with Gaussian noise and Poisson noise, respectively. (d)(e)(f) Noisy images captured with iPhone 11, Fujifilm X100T and Sony A7R2, respectively.}
  \label{real_3camera}
\end{figure} 

Capturing clean-noisy paired data for model training is time-consuming and inflexible for different cameras, so here we add noise to the high-quality NIR-RGB database to synthesize training data instead. A precise noise modeling is crucial for the final results. 
Many denoising algorithms assume the noise to be Gaussian or Poisson \cite{tassanoFastdvdnetRealtimeDeep2020}\cite{xueVideoEnhancementTaskoriented2019}. However, test results in Fig. \ref{real_3camera} show that these models can not describe the true low light noise well. Besides, low-light photos taken with different cameras vary a lot. These facts inspire us to use an integrated physical-process-based noise model being able to incorporate the various noise sources and compatible with different sensors. 

A typical CMOS camera sensor takes three steps to convert photon hit on the photosensor to the final digital images: the photosensor detects photons and converts them to electrons, integrated circuits convert and amplify electrons to voltage, and an ADC quantizes analog voltage signals to digital signals. During these stages, various sources of noise exist. We build an integrated noise model taking all the fundamental ones into account and adopt proper probabilistic distribution for each term to approximate the physical process best.

\vspace{2mm}
\noindent{\em (i) Shot noise.~~~~}
The shot noise is caused by the randomness of photon arrival and can be modeled with Poisson distribution with the mean value being the expected photon number $N$  \cite{konnikHighlevelNumericalSimulations2014}. The shot noise is determined by the signal itself and the photons at different color channels vary. The shot noise at channel $c$ is
\begin{equation}
    n^{\text{sht}}_{c}\sim P(N_c), c \in \{{\text{R, G, B, NIR}}\},
\end{equation}
with $P(\cdot)$ denoting the Poisson distribution. 

\vspace{2mm}
\noindent{\em (ii) Dark current.~~~~}
Dark current is caused by the random generation of electrons and holes, leading to a signal-independent noise \cite{konnikHighlevelNumericalSimulations2014}. The modeling of this noise is, however, more complex. In some prior works, dark current is assumed to follow a Gaussian or Poisson distribution \cite{wangEnhancingLowLight2019}\cite{konnikHighlevelNumericalSimulations2014}. However, researchers in \cite{weiPhysicsbasedNoiseFormation2020} observe the long-tail shape of noise data and propose to model this noise by Tukey lambda distribution, which matches the true dark current well. Inspired by this work, we model this noise using a clipped Poisson distribution with better approximation and compatibility with different sensors: 
\begin{equation}
  n^{\text{dk}} = \max\{0, n_d - N_d\},\ n_d \sim P(N_d).
\end{equation}
Here $N_d$ denotes the expected number of dark current electrons of each pixel.

\vspace{2mm}
\noindent{\em (iii) Read noise and Dynamic streak noise.~~~~}
During the conversion from electrons to voltage signals, read noise appears. This kind of noise does not depend on the signal and can be modeled with a Gaussian distribution and added to the signal directly as
\begin{equation}
n^{\text{rd}} \sim G(0, \sigma_r^2).
\end{equation}
In low light imaging, image quality would be further worsened by dynamic streak noise (DSN) \cite{wangEnhancingLowLight2019}, which appears to be horizontal streaks in low light noisy photos. This noise heavily depends on the pixel location. Combining shot
 noise, dark current, read noise and DSN, the equation would be
\begin{equation}
y_{r,c} = \beta_{r,c}(n^{{\text{sht}}}_c+n^{\text{dk}}+n^{{\text{rd}}})
\end{equation}
where $r$ denotes the row index of pixel and $\beta_{r,c} \sim G(1, \sigma_{\beta})$.

\begin{figure}[t!]
  \centering
  \includegraphics[width=\linewidth]{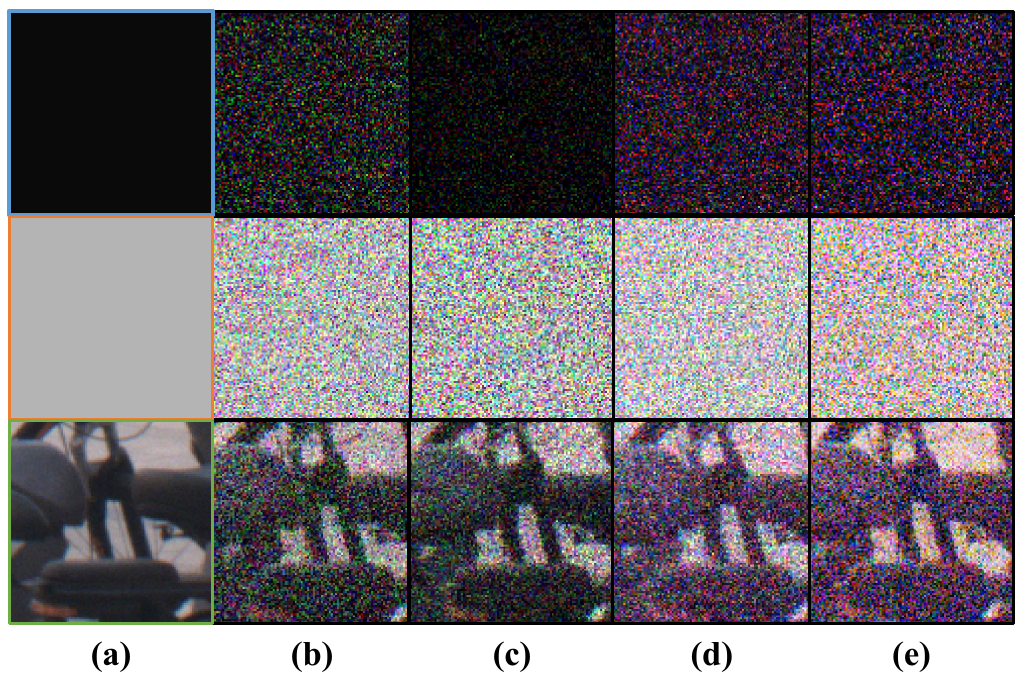}
  \caption{Comparison of noisy images produced by different noise models with the real images captured under low illumination. (a) Clean image patch for simulation. (b) Results by additive white Gaussian noise. (c) By Poisson noise. (d) By our physical-process-based noise model. (e) Real noisy image patch of the same scene.}
  \label{noise}
\end{figure}

\vspace{2mm}
\noindent{\em (iv) Quantization.~~~~}
Under low illumination, one usually set a high information gain to amplify the analog signal before quantization, and magnify further digitally for better visualization and subsequent processing. Mathematically, the voltage signals from camera sensor (with shot noise and dark current) are amplified $K_{a,c}$ times before digitalized, and the quantized signals can be expressed as
\begin{equation}
y_{q,c} = \lfloor K_{a,c}\beta_{r,c}(n^{{\text{sht}}}_c+n^{\text{dk}}+n^{{\text{rd}}}) \rfloor,
\end{equation}
where $\lfloor \cdot \rfloor$ denotes the floor function. Afterwards, the images can be amplified to fit a proper scale, e.g. [0, 255], with digital gain $K_d$.

\vspace{2mm}
In this paper, we integrated all above terms to model the noise based on the physical process of image recording. The model can be formulated as 
\begin{equation}
y_c=K_d (\lfloor K_{a,c}\beta_{r,c}(n^{{\text{sht}}}_c+n^{\text{dk}}+n^{{\text{rd}}}) \rfloor), \label{cmosmodel}
\end{equation}
in which the parameters of noise terms
$n^{{\text{sht}}}_c$, $n^{\text{dk}}$, and $n^{{\text{rd}}}$
are all defined and calibrated in a pixel-wise manner.


\vspace{3mm}
To synthesize noisy low light images, we set parameter $N_c$ based on the target flux and clean image pixel and simulate system gain $K = K_a K_d$. 
Suppose we capture videos in extremely dark environments, and set the highest analog amplification factor $K_{a,c}$ and a digital gain adjust $K_d=\frac{K}{K_a}$ is set. Then the number of photons $N_c$ can be infered as $N_c = \frac{{\bf{\text I}}_c}{K}$ to maintain the overall brightness.
Parameters are calibrated following the procedure described in \cite{wangEnhancingLowLight2019}, and then randomly fluctuates within a range to improve robustness. The ranges of the key parameters are shown in Tab. \ref{model_para}. Note that image noise intensity is mainly determined by the system gain $K$. 
Fig. \ref{noise} demonstrates the simulated noisy data and comparison with results by other models. We can see that our noise model provides simulation results much closer to real captured low light video frames, especially in the dark regions.


\section{Image Acquisition}
\subsection{The RGB-NIR Imaging System}

\begin{figure}[t]
  \centering
  \includegraphics[width=\linewidth]{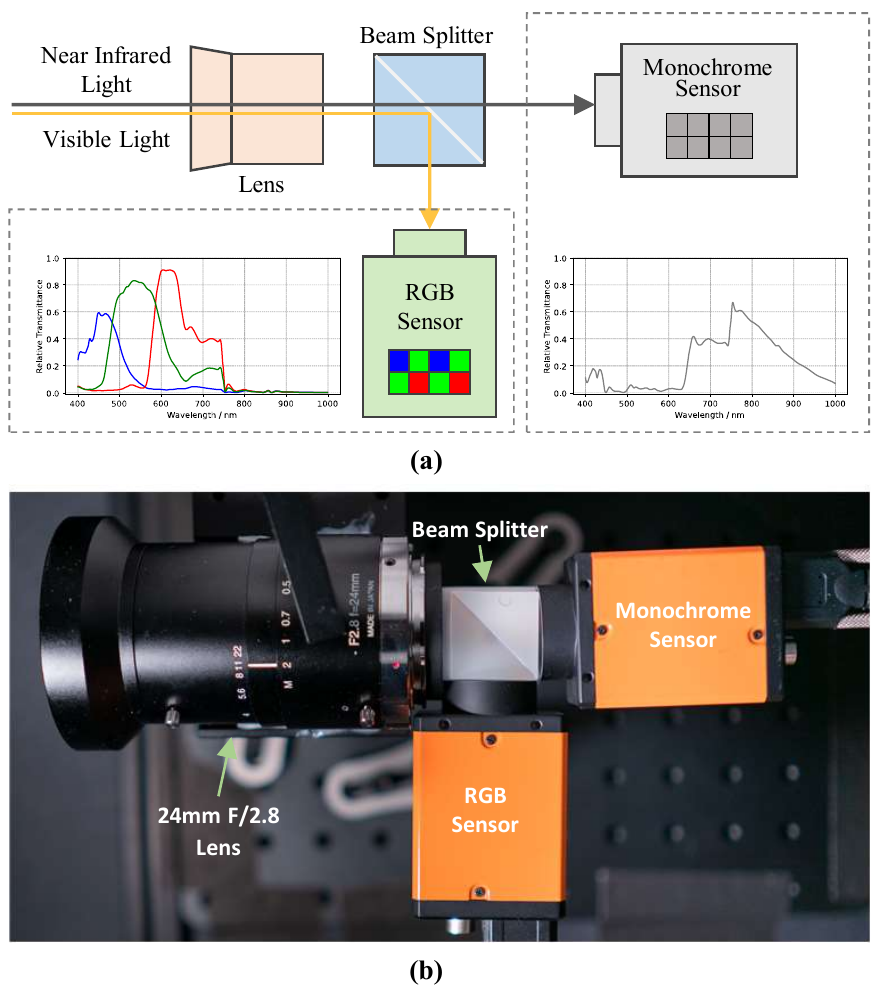}
  \caption{Our RGB-NIR imaging system. \textbf{(a)} Diagram of our two sensor acquisition system, with the sensor response of the visible band and near-infrared plotted. \textbf{(b)} A photo of our prototype.
  }
  \label{fig:system}
\end{figure}
To capture paired RGB and NIR videos, we build an RGB+NIR camera with two 1" CMOS image sensors (HIKROBOT MV-CH089-UC and MV-CH089-10UM) recording the visible and near-infrared bands respectively. As illustrated in Fig.~\ref{fig:system}(a), the target scene is firstly captured by the primary lens (Nikon BlueVision BV-L1024, $f$=24mm, F-Mount), and then the outgoing light is split by a cubic beam splitter which reflects RGB light (stops at roughly 700nm) and transmits NIR light. The spectral transmission curve of the beam splitter and simulated spectral response of red, green, blue, and near-infrared channels are also plotted in Fig. \ref{fig:system}(a). 
Later, two arms arrive at two orthogonally placed sensors: the RGB sensor UC with Bayer CFA captures RGB images, and monochrome sensor UM without a near-infrared cut-off filter captures monochromatic images). 
These two arms are complementary in collecting photons falling within the broad spectral region (350-1100 nm) of two CMOS sensors. 
Two cameras are wired to a PC and are software synchronized.
After capturing the raw video pairs, we need to align the images from two sensors. Since the RGB and NIR images share the same optical axis, there is no parallax and a precise alignment can be achieved with a simple SIFT detector \cite{loweObjectRecognitionLocal1999} and affine transformation. The pixel resolution of the aligned video pair is around $1280\times 720$.

\begin{figure}[t!]
  \centering
  \includegraphics[width=\linewidth]{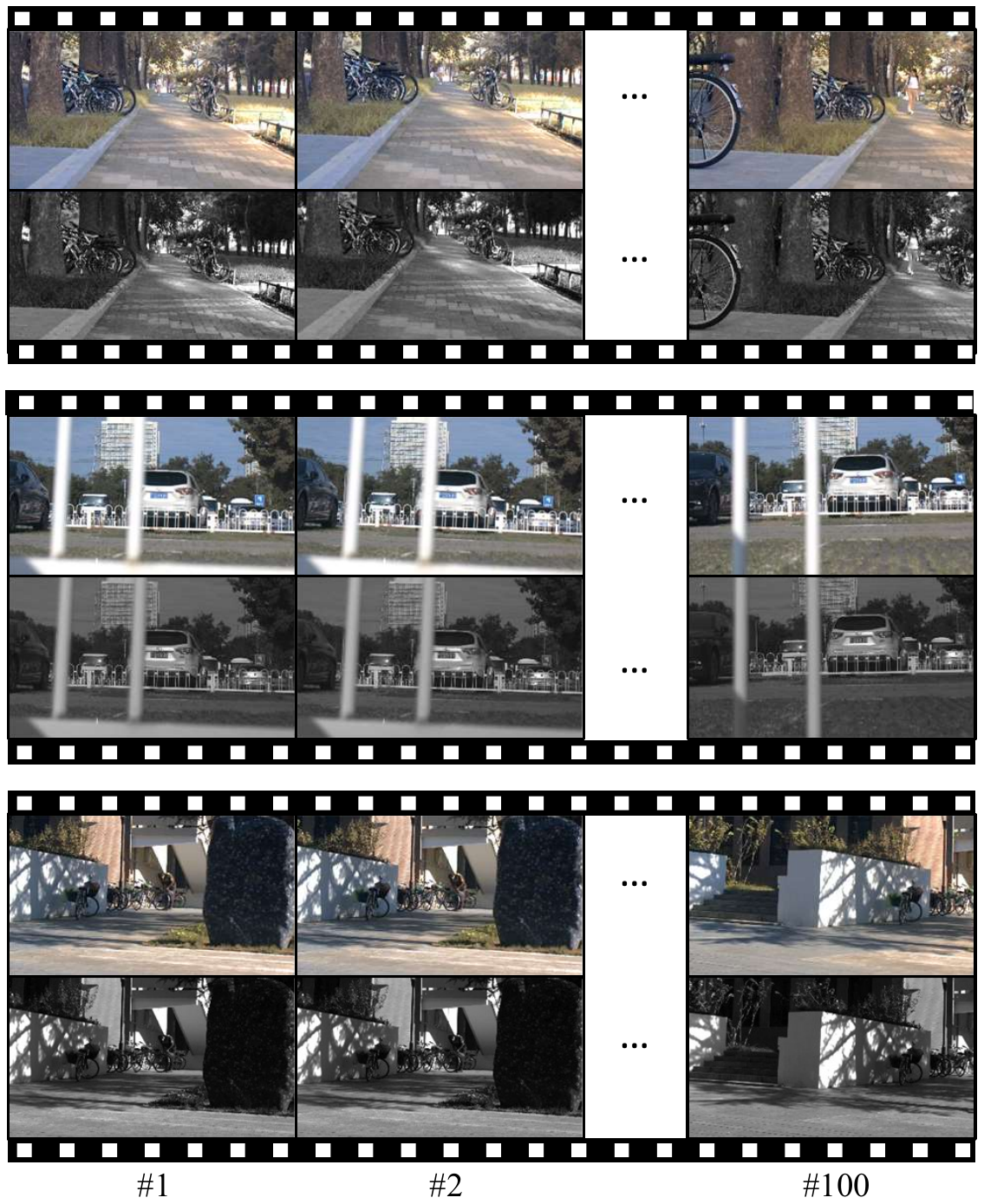}
  \caption{Several examples from our RGB-NIR video dataset, with pixel-wise registration bewteen measurements in two wavelength bands.}
  \label{dataset}
\end{figure}


\begin{table*}[t!]
  \renewcommand{\arraystretch}{1.8}
  \small
  \centering
    \caption{Comparison of performance in terms of PSNR (dB) / SSIM on our RGB-NIR video dataset. $K=K_a K_d$ denotes the total gain factor during the simulation process. S: Spatial, T: Temporal, C: Spectral. ``Ours S'' denotes the results from SCMAN\_Fn1, with only spatial prior. ``Ours S+'' denotes the results from SCMAN\_Fn7, with explicit exploration of temporal prior. ``Ours S+T+C'' denotes the results from DCMAN\_Fn7 (or DCMAN), with spatio-temporal and spectral prior jointly.  "Ours$-$LSTM" denotes the results by removing bi-ConvLSTM module from DCMAN\_Fn7.}
  \begin{tabular}{c|ccc|cccc}
    \hline
    Simu. & V-BM4D \cite{maggioniVideoDenoisingDeblocking2012}  & TOFlow \cite{xueVideoEnhancementTaskoriented2019} & FastDVDnet \cite{tassanoFastdvdnetRealtimeDeep2020}  & Ours S & Ours S+T & Ours$-$LSTM & Ours S+T+C\\ \hline
    $K=10$ & 27.974 / 0.867 & 28.976 / 0.914 & 29.020 / 0.919 & 29.530 / 0.906 & 31.062 / 0.935 & 31.121 / 0.936 & {\bf 31.150 / 0.937} \\
    $K=20$ & 24.017 / 0.790 & 25.943 / 0.873 & 26.012 / 0.883 & 28.127 / 0.879 & 29.838 / 0.916 & 29.863 / 0.918 & {\bf 29.965 / 0.920} \\
    $K=40$ & 19.667 / 0.659 & 22.227 / 0.802 & 22.371 / 0.819 & 26.442 / 0.841 & 27.947 / 0.882 & 28.256 / 0.889 & {\bf 28.427 / 0.892} \\ \hline
  \end{tabular} 
  \label{tab:simu_sota}
\end{table*}


\begin{figure*}[t!]
  \centering
  \includegraphics[width=\textwidth]{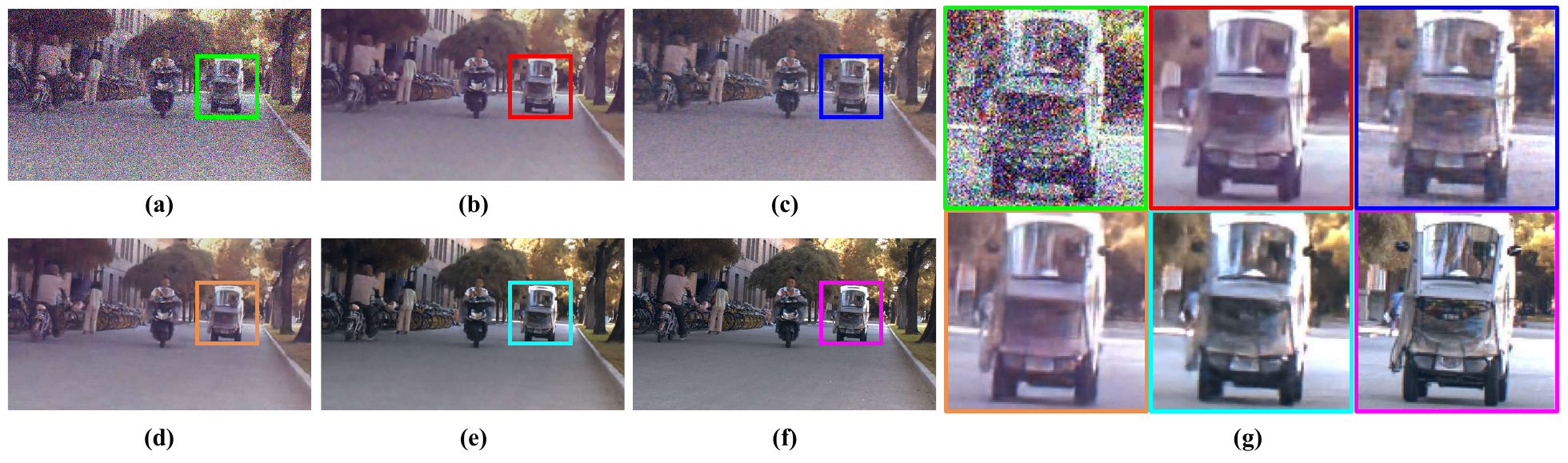}
  \caption{Simulation test results of our model compared with state of the art methods V-BM4D \cite{maggioniVideoDenoisingDeblocking2012}, TOFlow \cite{xueVideoEnhancementTaskoriented2019}, and FastDVDnet \cite{tassanoFastdvdnetRealtimeDeep2020}, and the degenerated single-channel version of our deep network (SCMAN), i.e.,model without NIR channel. (a) Simulated noisy frame. (b) V-BM4D. (c) TOFlow. (d) FastDVDnet. (e) Ours. (f) Groundtruth. In (g) we show the comparison in the highlighted region for better visualization, with the same layout as (a)-(f). 
 }
  \label{fig:simu}
\end{figure*}

\subsection{Data for Model Training and Testing}

For simulating the clean-noisy data for supervised model training and quantitative performance evaluation, we collect a dataset including 118 high-quality RGB-NIR video clips with 11,444 frames of real nature scenes under sufficient illuminance. Using a high-end commercial camera as a reference, we adjust the color tone and image contrast by building a Look-Up-Table (LUT) to improve the quality of our dataset. Several examples are shown in Fig.~\ref{dataset}. The training data for the network is generated by adding noise to the bright dataset following the aforementioned CMOS noise model. For quantitative evaluation, 104 videos are used for model training and the rest for testing.

To test the performance on real low light videos, we captured noisy input in dark environments with around 0.125\%$\sim$0.5\% illuminance of daily bright photography. 
These testing data are preprocessed in the same way as the dataset and linearly scaled to normal brightness before being fed into the trained deep network. 


\section{Experiments and Analysis}

\subsection{Experiment Settings}
The training set for the DCMAN is synthesized by adding noise to our RGB-NIR clean video pairs following the CMOS noise model, with the parameters listed in Table.~\ref{model_para}. To raise the robustness towards different noise levels, parameters $K_a$ and $K_d$ are randomly selected for each simulated low light video.

\begin{table}[ht]
  \renewcommand{\arraystretch}{1.8}
  \small
  \centering
  \vspace{3mm}
\caption{Parameter settings of our CMOS noise model for generating paired training data. Each parameter is randomly selected from a uniform distribution over a range instead of a fixed value, which can help cover the intensity variation among different scenes.}
\begin{tabular}{l|l}
    \hline
    \textbf{ ~~~~~~~~~~~~Parameters ~~~~~~~~~~~~} & ~~~~~~~\textbf{Distribution}~~~~~~~ \\ \hline
    $N_d$: Dark current electron number & 
    ~~$U(2, 10)$ \\ \hline
    $\sigma_{\beta}$: Standard deviation of DSN & ~~$U(0.02,0.08)$ \\ \hline
    $\sigma_{r}$: Standard deviation of read noise & ~~$U(0.5,2)$ \\ \hline
    $K_{RGB}$: Gain in RGB channel& 
    ~~$U(10,40)$\\ \hline
    $K_{NIR}$: Gain in NIR channel & ~~$U(K_{RGB},3K_{RGB})$ \\ \hline
  \end{tabular}
  \label{model_para}
\end{table}

We conduct experiments on a PC with Intel Core CPUs and NVIDIA GeForce GTX 1080Ti GPUs. The DCMAN network is trained with 20,000 paired video patches randomly cropped from original synthesized videos with a patch size of $120\times 120$ pixels. We use the Adam optimizer to train the network. The batch size is 10, the learning rate is $10^{-4}$ and decreases by a factor of 0.1 every epoch. The color loss parameter $\lambda_2$ is 0.1 initially and increases to 0.25 and 0.6 after the 20th and 30th epoch, respectively, and $\lambda_1 = 1 - \lambda_2$. We terminate the training after 40 epochs. The whole training process takes around 40 hours.

As the first to discuss low light video reconstruction with RGB and NIR video pairs, in order to prove our superior performance towards RGB video methods, our method is compared with three state-of-the-art video denoising methods, one filtering based and two data-driven methods: V-BM4D \cite{maggioniVideoDenoisingDeblocking2012}, FastDVDnet \cite{tassanoFastdvdnetRealtimeDeep2020} and TOFlow \cite{xueVideoEnhancementTaskoriented2019}. 
For a fair comparison, we conduct the following preprocessing to optimize the performance of three benchmark methods. 
For V-BM4D, information on noise level is required and we use its in-built noise estimator to estimate the sigma value (standard variation of the Gaussian noise). For FastDVDnet, we remove the noise map and re-train the model on RGB videos in our RGB-NIR dataset to maintain fairness. TOFlow requires the optical flow between adjacent frames, so we use the pre-trained flow estimation network SpyNet \cite{ranjanOpticalFlowEstimation2017} and fine-tune it on our dataset with superimposed Gaussian noise, as mentioned in its original paper.

\begin{figure*}[t]
  \centering
  \includegraphics[width=\linewidth]{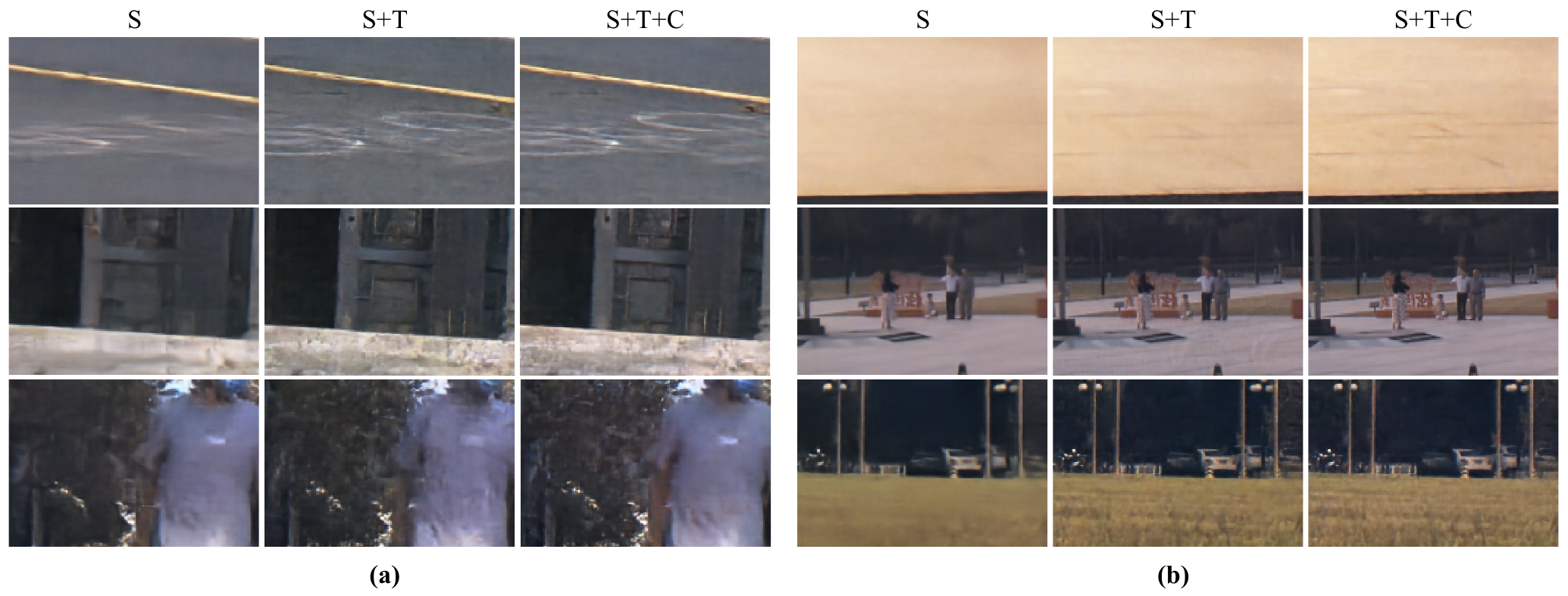}
  \caption{Comparison of SCMAN\_Fn1 ({\bf S}patial information), SCMAN\_Fn7 ({\bf S}patial+{\bf T}emporal information) and DCMAN\_Fn7 ({\bf S}patial+{\bf T}emporal+Spe{\bf c}tral information) results to verify the effectiveness of our multi-frame dual-channel network. (a) and (b) show the comparion on synthetic and real data, respectively.
  }.
  \label{fig:dcsc}
\end{figure*}

\begin{table*}[t!]
  \renewcommand{\arraystretch}{1.6}
  \small
  \centering
  \caption{Performance on simulation test for several $F_n$ and system gain $K$. Analytical data are shown as $PSNR_{DCMAN} / PSNR_{SCMAN} (dB)$. $PSNR_{DCMAN}$ denotes PSNR of DCMAN (with NIR channel), and $PSNR_{SCMAN}$ denotes that of SCMAN (without NIR channel).}  
  \begin{tabular}{ccccc}
    \hline
    ~~~~~~~~Simultion Test~~~~~~~~ &~~~~~~~~~~ $F_n=1$ ~~~~~~~~~~ &~~~~~~~~~~ $F_n=3$ ~~~~~~~~~~ &~~~~~~~~~~ $F_n=5$ ~~~~~~~~~~& ~~~~~~~~~~$F_n=7$~~~~~~~~~~ \\ \hline
    $K=10$ & 29.851 / 29.530  & 30.826 / 30.603 & 31.166 / 30.889 & 31.150 / 31.062  \\
    $K=20$ & 28.515 / 28.127  & 29.521 / 29.192 & 29.898 / 29.533 & 29.965 / 29.838  \\
    $K=40$ & 26.875 / 26.442  & 27.875 / 27.475 & 28.275 / 27.847 & 28.427 / 27.947  \\ \hline
  \end{tabular} 
  \label{tab:temporal_spectral}
\end{table*}

\subsection{Experiments on Synthetic Noisy Videos}

To quantify our performance, we first apply our methods on the simulated test dataset, using PSNR (Peak signal-to-noise ratio) and SSIM (The Structural Similarity Index) as evaluation metrics. We also compare with three state-of-the-art methods, as shown in Tab. \ref{tab:simu_sota}, including a patch-based video denoising method V-BM4D \cite{maggioniVideoDenoisingDeblocking2012}, a deep-learning-based method FastDVDnet \cite{tassanoFastdvdnetRealtimeDeep2020}, and an optic-flow-based deep learning method TOFlow \cite{xueVideoEnhancementTaskoriented2019}. 

The experiment is performed under the setting $K_{RGB} = K_{NIR}$. Other noise parameters are set to match Tab.~\ref{model_para}. The results show that our method outperforms other methods by a big gap. At $K=40$, the PSNR increment goes up to 8.76 dB, 6.20 dB, and 6.06 dB higher than V-BM4D, TOFlow, and FastDVDnet, respectively. The visual comparison in Fig.~\ref{fig:simu} displays similar advantages. One can see that V-BM4D and TOFlow leave some noise and FastDVDnet tends to produce oversmooth results, while the proposed approach preserves much more details and better color fidelity. 

The superior performance benefits from multiple aspects: 
firstly, more information from the infrared channel is utilized in our methods, while V-BM4D and FastDVDnet only use RGB input. Secondly, 7 adjacent frames are used to denoise the center frame instead of 5 in FastDVDnet.  Finally, two previous deep-learning-based methods (TOFlow and FastDVDnet) are retrained on our data but both assume Gaussian noise, while our models are trained with data synthesized following an integrated CMOS noise model and calibrated parameters. 

\subsection{Ablation Studies}

\noindent{\bf Temporal prior information.~~~~}
In our network, the LSTM module combines information from  $F_n = 2T + 1$ frames to denoise the center frame. To evaluate the influence from temporal information utilization, we set $F_n$ to be 1, 3, 5, and 7 respectively to train 4 models DCMAN\_Fn1, DCMAN\_Fn3, DCMAN\_Fn5, and  DCMAN\_Fn7. Their performances are shown in Tab.~\ref{tab:temporal_spectral}. Generally, the denoise performance grows as $F_n$ increases. 
Empirically, in our final network, we set $F_n = 7$, which provides the network with a massive amount of temporal prior assistance. Comparing DCMAN\_Fn1 with DCMAN\_Fn7 ($K=40$), i.e., the ``Ours S'' and ``Ours S+T'' columns in Tab. \ref{tab:simu_sota}, introducing temporal information increases PSNR by 1.55 dB. 


\vspace{2mm}
\noindent{\bf Spectral prior information.~~~~}
One of the key contributions of our work is the utilization of additional NIR information. 
To quantitatively evaluate the performance improvement brought by the NIR information, we disable the NIR channel in our network and remove the corresponding input to train a network recovering only RGB video frames. The network is called Single-Channel Multi-frame Attention Network (SCMAN), as shown in the two parallel insets of Fig.~\ref{net}. 

We firstly vary the frame number ($F_n=1,3,5,7$) to train SCMAN\_Fn1, SCMAN\_Fn3, SCMAN\_Fn5, and SCMAN\_Fn7, with the results shown in Tab.~\ref{tab:temporal_spectral}. We can see that introducing NIR information indeed raises the performance at almost all of the settings. 
Comparing SCMAN\_Fn7 with DCMAN\_Fn7 ($K=40$), we can calculate that spectral information increases PSNR by 0.48 dB on our test dataset, as shown in the ``Ours S+T+C'' column in Tab. \ref{tab:simu_sota}. We also compare the reconstruction results from DCMAN\_Fn7 and SCMAN\_Fn7 visually in Fig. \ref{fig:dcsc}. The conclusion consists well with the quantitative results, more details and fewer artifacts occur in DCMAN results than in SCMAN.

\vspace{2mm}
\noindent{\bf Conv-LSTM.~~~~}
Bidirectional Conv-LSTM modules are used to encode temporal correlation among multiple frames. However, features from different frames may also be combined directly with a multi-channel CNN, named CNN early fusion.
To quantify the contribution from the Conv-LSTM module, we design a new network with CNN early fusion \cite{caballeroRealtimeVideoSuperresolution2017}.
The result with Conv-LSTM and trivial CNN early fusion are shown in the `Ours S+T+C' and `'Ours$-$LSTM' columns in Tab. \ref{tab:simu_sota}. It is observed that on our data the Bi-ConvLSTM module increases PSNR by 0.17.


\vspace{2mm}
\noindent{\bf The design of the deep network.~~~~}
The superior performance of our final model can be attributed to both better network design, additional information from the NIR channel, and adopting a physical-process-based noise model.  
Considering that most state-of-the-art denoising methods are trained with Gaussian or other simple noise models, we conduct an experiment here to verify the advantages of the proposed network itself. 
Specifically, we train a DCMAN and SCMAN model on additive white Gaussian noise (AWGN) and name it DCMAN\_G and SCMAN\_G. Compared to state-of-the-art deep learning-based video denoising methods, TOFlow \cite{xueVideoEnhancementTaskoriented2019} and FastDVDnet \cite{tassanoFastdvdnetRealtimeDeep2020}, retrained with synthetic data by adding AWGN on our dataset, both our single-channel and bi-channel networks perform much better, as shown in Tab.~\ref{gauss_simu}. 

\begin{table}[ht]
  \renewcommand{\arraystretch}{1.5}
  \small
  \centering
  \caption{Test results on image synthesized with Additive White Gaussian Noise (AWGN) compared with two deep learning based method TOFlow and FastDVDnet. DCMAN\_G and SCMAN\_G are our methods with and without NIR channel trained on simulated AWGN noise. Our model outperform those methods on both noise levels.}  
  \begin{tabular}{ccc}
    \hline
    PSNR(dB) / SSIM  &~~~~~~~~$\sigma=20$~~~~~~~~&~~~~~~~~$\sigma=40$~~~~~~~~\\ \hline
    TOFlow & 31.792 / 0.941 & 29.328 / 0.902 \\ \hline
    FastDVDnet & 32.509 / 0.947 & 29.605 / 0.911 \\ \hline
    SCMAN\_G & 32.833 / 0.952 & 30.259 / 0.923\\ \hline
    DCMAN\_G & {\bf 32.988 / 0.954} &  {\bf 30.418 / 0.926} \\ \hline
  \end{tabular}
  \label{gauss_simu}
\end{table}

\begin{figure}[ht]
  \centering
  \includegraphics[width=\linewidth]{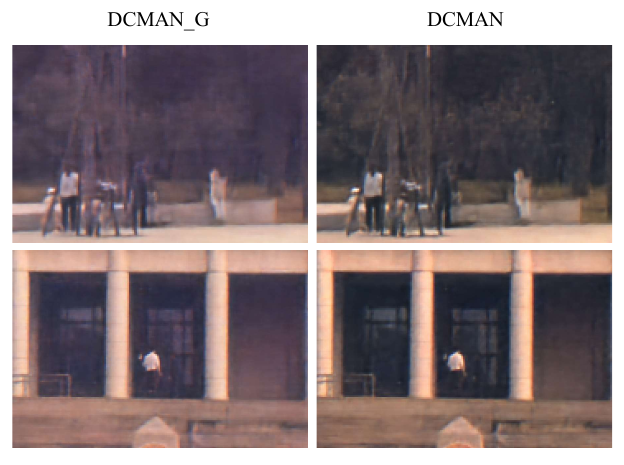}
  \caption{Comparison of results by DCMAN\_G (trained from data with Gaussian noise) and DCMAN (trained with synthetic data following physical CMOS noise model) on real low light videos.}
  \label{real_gauss}
\end{figure}

\vspace{2mm}
\noindent{\bf Integrated noise model.~~~~}
Our training data are synthesized by a physical-process-based integrated noise model. 
To validate the necessity of using such a noise model incorporating various noise sources, we compare the performance of DCMAN\_G (Gaussian noise) and DCMAN (calibrated noise following the integrated noise) on real low light images. The results are shown in Fig. \ref{real_gauss}, from which one can see noticeable performance 
improvements brought by learning from data generated by an integrated noise model with calibrated model parameters. 

\subsection{Experiments on Real Captured Low Light Videos}

\begin{figure*}[t!]
  \centering
  \includegraphics[width=\textwidth]{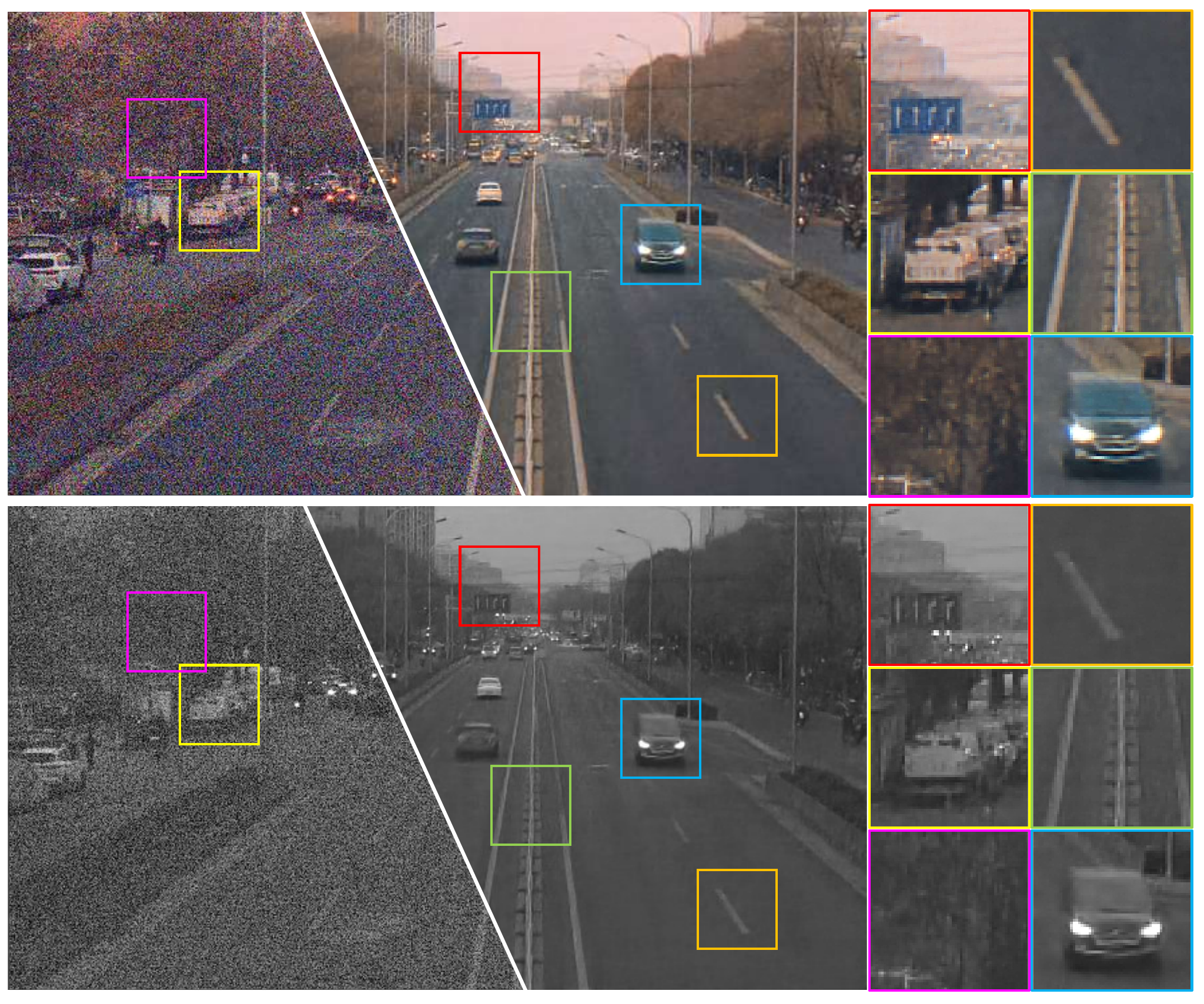}
  \caption{Our RGB and NIR reconstruction results on a dark video frame captured by our setup at night.}
  \label{fig:real_big}
\end{figure*}

\begin{figure*}[t!]
  \centering
  \includegraphics[width=\textwidth]{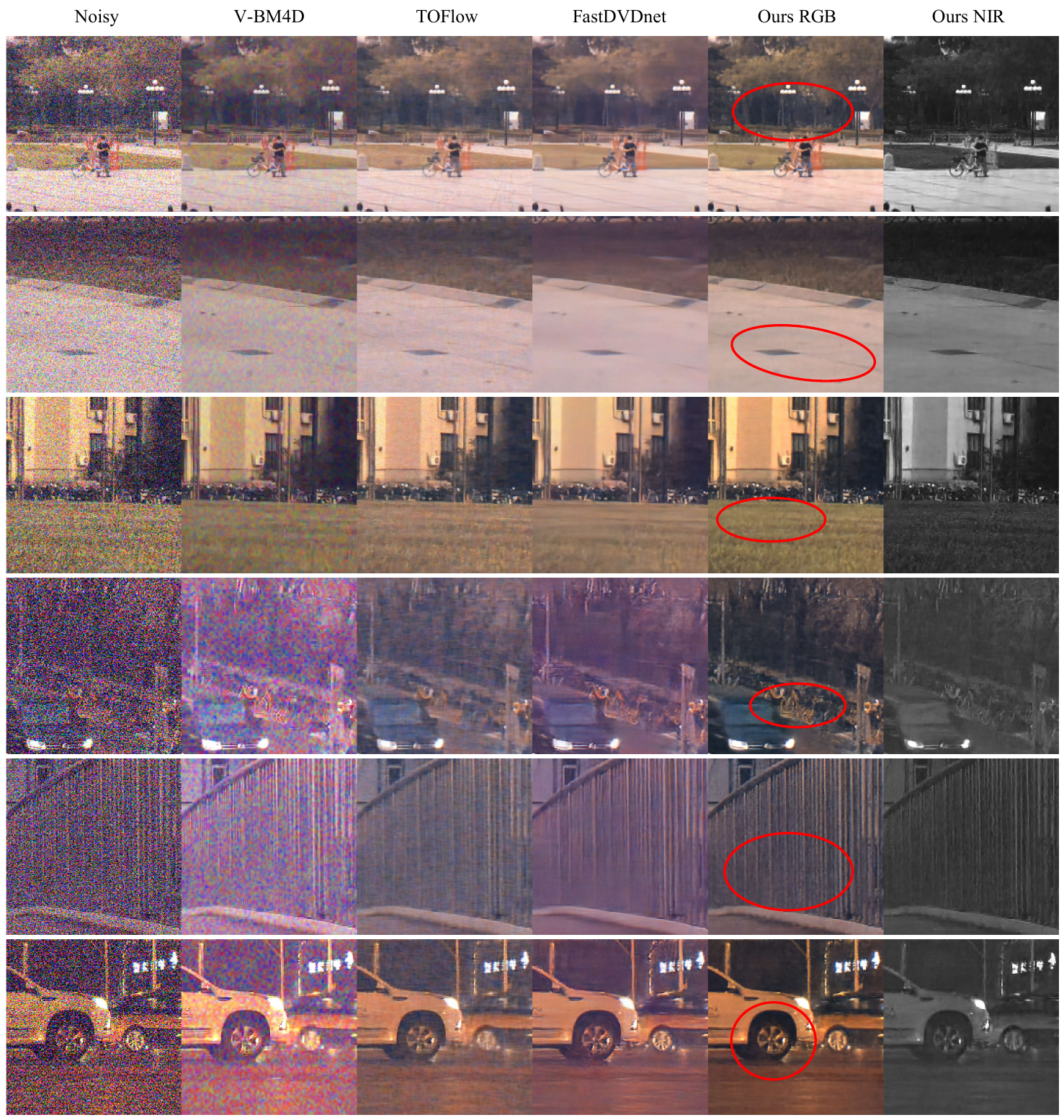}
  \caption{Our RGB reconstruction results on real low light video frames and comparison with state-of-the-arts. Here some regions with large performance diversity are highlighted with red ellipses.}
  \label{fig:real}
\end{figure*}

We test our approach on real low-light videos captured with our dual sensor camera, the results are shown in Fig. \ref{fig:real_big}-\ref{fig:real}, and please refer to supplementary data for video demonstration. The images are captured at dark hours with illumination of less than 1 lux. From the image in Fig.~\ref{fig:real_big}, one can see that under such low illuminations the input raw images are severely deteriorated by noise, with structure details washed out and color deviating largely. After reconstruction, we can achieve striking quality improvement. The flat regions (e.g., road, sky), structures (e.g., road lamp, driving line, car, traffic sign), and textured regions (e.g., trees, shrubbery) are all recovered with high quality. The high performance indicates that our approach is quite promising in assisting night traffic surveillance and auto driving.

Comparatively, three previously published state-of-the-arts show limited performance in such challenging scenarios, as shown in Fig.~\ref{fig:real}. The results from V-BM4D and TOFlow tend to suffer from residual noise, while FastDVDnet causes over-smoothness and cannot recover thin structures. Besides, all these three methods have apparent color distortion, while our DCMAN removes purplish color bias and preserves image details well. For example, 
in the 1st row, we can notice that the details in the dark background are recovered, such as the tree trunks and the shrubs. In the 2nd and 4th scenes, the details are reconstructed at a much higher quality, e.g., the textures on the marble, and the bike on the side of the road. Besides, the proposed approach produces high performance on the dense striped patterns (the railing in the 5th row) that are prone to oversmoothness and the highlight region (car wheel in the 6th row). Our results are also advantageous in the texture regions, with more details and less color distortion, such as the lawn regions in the 3rd scene. Overall, the proposed method is of stronger noise suppression, better structure preservation, and higher color fidelity. The superior performance is mainly attributed to the additional information from the NIR channel and successful exploration of the cross-channel and multi-frame prior, including both the network design and effective training strategy. 


\section{Summary and Discussions}
This paper reports a dual-channel computational dark videography approach with superior performance than state-of-the-arts, by collecting more photons optically and introducing cross channel priors computationally. The high performance benefits from technical contributions in three folds: (i) We design a compact RGB+NIR dual-sensor camera to largely increase the collected photons of a conventional RGB camera by additionally capturing a NIR wavelength band, and two paths are with pixel-wise registration and high precision synchronization.  (ii)  We then propose a Dual-Channel Multi-frame Attention Network (DCMAN) to enhance the RGB and NIR video pair in an end-to-end manner by exploring the spatial, temporal, and spectral information jointly. (iii) The model can be learned effectively from synthetic data produced by superimposing noise on our high quality NIR+RGB dataset following an integrated physical-process-based noise model and adapted to different sensors. 

So far, the application of our approach is somewhat limited due to its high computing complexity. For a 360 $\times$ 640 video frame, we need around 700ms to recover the high-quality video with an NVIDIA GeForce GTX 1080Ti GPU. A low weight model is under development currently. 
In the hardware, one can replace the Bayer pattern filter with a customized one with NIR transmittance to collect more photons in dark environments, and developing corresponding reconstruction algorithms is a worth studying topic.
The approach can also be extended further to other wavelength bands for different imaging scenarios or tasks, such as UV, mid-NIR, etc. 



%

\ifCLASSOPTIONcompsoc
  \section*{Acknowledgments}
\else
  \section*{Acknowledgment}
\fi

Our code will be soon available at https://github.com/ jarrycyx/dual-channel-low-light-video.git

\ifCLASSOPTIONcaptionsoff
  \newpage
\fi


\bibliographystyle{bib/IEEEtran}
\bibliography{DCMAN}

%

\begin{IEEEbiography}[{\includegraphics[width=1in,height=1.25in,clip,keepaspectratio]{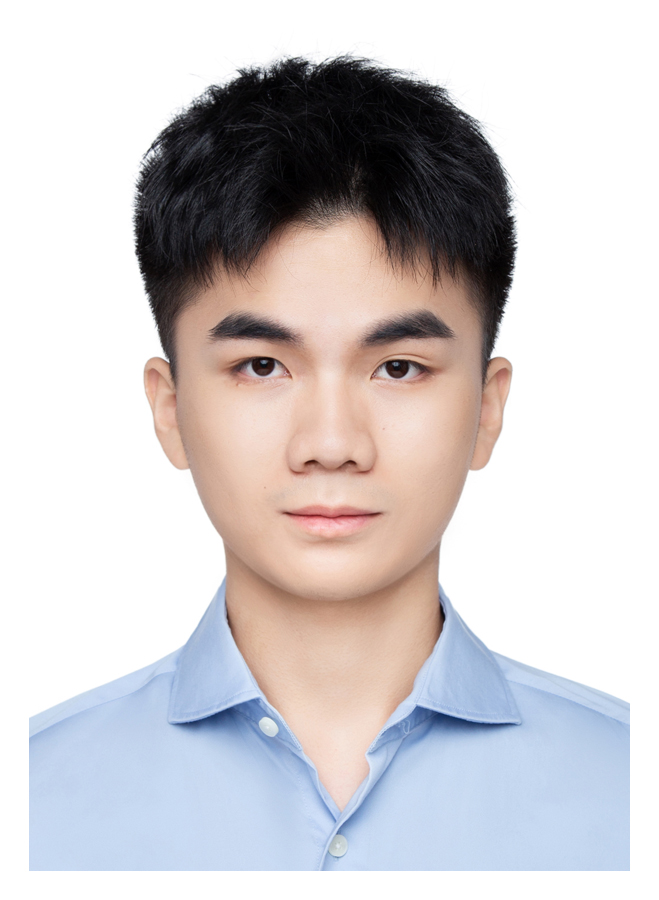}}]{Yuxiao Cheng}
is an undergraduate student in the Department of Automation, Tsinghua University, Beijing, China. His research interests include computational imaging and computer vision.
\end{IEEEbiography}

\begin{IEEEbiography}[{\includegraphics[width=1in,height=1.25in,clip,keepaspectratio]{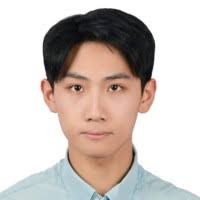}}]{Runzhao Yang}
received his B.E. degree in the School of Electrical Engineering and Automation from Wuhan University, Wuhan, China, in 2020. He is pursuing his Ph.D. degree in the Department of Automation at Tsinghua University, Beijing, China. His research interests include computer vision, data compression, and machine learning.
\end{IEEEbiography}

\begin{IEEEbiography}[{\includegraphics[width=1in,height=1.25in,clip,keepaspectratio]{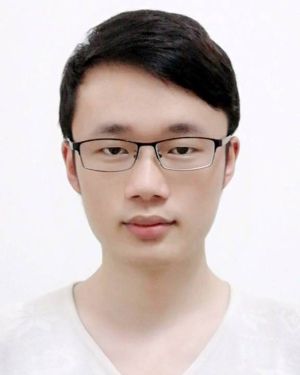}}]{Zhihong Zhang}
  received the B.Eng. degree from the School of Electronics Engineering, Xidian University, Xi'an, China, in 2019. He is currently pursuing the Ph.D. degree in the Department of Automation, Tsinghua University. His research interests include computational imaging, computer vision, and machine learning.
\end{IEEEbiography}


\begin{IEEEbiography}[{\includegraphics[width=1in,height=1.25in,clip,keepaspectratio]{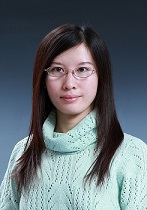}}]{Jinli Suo}
received the BS degree in computer science from Shandong University, Shandong, China, in 2004 and the Ph.D. degree from the Graduate University of Chinese Academy of Sciences, Beijing, China, in 2010. She is currently an associate professor with the Department of Automation, Tsinghua University, Beijing, China. Her research interests include computer vision, computational photography, and statistical learning. She is an Associate Editor for the IEEE Transactions on Computational Imaging.
\end{IEEEbiography}

\begin{IEEEbiography}[{\includegraphics[width=1in,height=1.25in,clip,keepaspectratio]{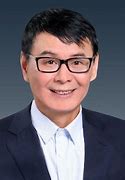}}]{Qionghai Dai}
received the M.S. and Ph.D. degrees in computer science and automation from Northeastern University, Shenyang, China, in 1994 and 1996, respectively. He is currently a Professor with the Department of Automation, an Adjunct Professor with the School of Life Sciences, Tsinghua University, and an Academician with the Chinese Academy of Engineering. He has authored or coauthored more than 200 conference and journal papers and two books. His research interests include computational photography and microscopy, computer vision and graphics, and intelligent signal processing. He is an Associate Editor for the Journal of Visual Communication and Image Representation, the IEEE Transactions on Neural Networks and Learning Systems, and the IEEE Transactions on Image Processing.
\end{IEEEbiography}




  
  

\end{document}